\documentclass{aa}

\usepackage{graphicx}
\usepackage{txfonts}
\usepackage{lipsum}
\usepackage{subcaption}         
\usepackage{lscape}             
\usepackage{placeins}           
                                
\usepackage{hyperref}
\usepackage{color,multicol}
\usepackage{tabularx}  
\usepackage[normalem]{ulem}
\usepackage{flushend} 

\def\kms{km~s$^{-1}$}
\def\msun{M$_{\sun}$}

\def\halpha{\ifmmode {\rm H{\alpha}} \else $\rm H{\alpha}$\fi}
\def\hbeta{\ifmmode {\rm H{\beta}} \else $\rm H{\beta}$\fi}
\def\oii{[\ion{O}{ii}]}
\def\oiid{[\ion{O}{ii}]$\lambda\lambda$3726,3729}

\begin{document}

    \title{Probing the molecular gas content of galaxies in an over-dense group at z$\sim$0.7: A test case for environmental quenching}
	\titlerunning{Probing the molecular gas content of galaxies in an over-dense group at z$\sim$0.7}

   \author{J. Freundlich\inst{1}
	\and
	B. Epinat\inst{2,3,4}
    \and
	T. Contini\inst{5}
	\and
	P. Salomé\inst{6}
    \and
    F. Combes\inst{6,7}
    \and
    B. Jego\inst{1}
    \and
    D. Krajnovi\'c\inst{8}
	\and
	W. Mercier\inst{2}
    \and
    C. Muñoz López\inst{8}
    \and
    M. Béthermin\inst{1}
    \and
	L. A. Boogaard\inst{9}
	\and
    A. Boselli\inst{2}
    \and
    R. Herrera-Camus\inst{10,11}
    \and
    D. Ismail\inst{1}
    \and
	F. Jiang\inst{12}
    \and
    K. Kraljic\inst{1}
    \and
    F. Renaud\inst{1}
	\and
	S. Tacchella\inst{13, 14}
}

\institute{
	Université de Strasbourg, CNRS, Observatoire astronomique de Strasbourg, UMR 7550, F-67000 Strasbourg, France\\ \email{jonathan.freundlich@astro.unistra.fr}
	\and
	Aix Marseille Univ, CNRS, CNES, LAM, Marseille, France
	\and
	French-Chilean Laboratory for Astronomy, IRL 3386, CNRS and Universidad de Concepción, Departamento de Astronomía, Barrio Universitario s/n, Concepción, Chile  
	\and
    Canada-France-Hawaii Telescope, 65-1238 Mamalahoa Highway, Kamuela, HI 96743, USA
	\and
    Université de Toulouse, CNES, CNRS, IRAP, Toulouse, France
    \and
	Observatoire de Paris, LUX, PSL University, Sorbonne University, CNRS, Paris, France
    \and
    Collège de France, Paris, France
    \and
    Leibniz-Institut für Astrophysik Potsdam (AIP), 14482 Potsdam, Germany
	\and
    Leiden Observatory, Leiden University, NL-2300 RA Leiden, The Netherlands
    \and
    Departamento de Astronomía, Universidad de Concepción, Concepción, Chile
    \and
    Millennium Nucleus for Galaxies, Concepción, Chile.
	\and
    Kavli Institute for Astronomy and Astrophysics, Peking University, China
	\and
	The Kavli Institute for Cosmology (KICC), University of Cambridge, Cambridge, CB3 0HA, UK
	\and
	Cavendish Laboratory, University of Cambridge, Cambridge, CB3 0HE, UK
}

   \date{Received June 24, 2026; Accepted September 10, 2026}

 
  \abstract 
   {
   We investigated the impact of group environment on molecular gas reservoirs at an intermediate redshift by observing the CO(2-1) emission from the galaxy group COSMOS-Gr30 at $z \sim 0.7$ with the Institut de Radioastronomie Millimetrique (IRAM) NOrthern Extended Millimeter Array (NOEMA) and 30m telescope. This dense environment, located at the intersection of large-scale cosmic web filaments hosts a large ($\sim 10^{4}$ kpc$^{2}$) ionized gas structure revealed by MUSE. We detect CO emission in four galaxies in the group with a signal-to-noise ratio (S/N) greater than five and derive upper limits for the remaining group members with secure spectroscopic redshifts. Stacked measurements indicate that group galaxies exhibit, on average, molecular gas contents reduced by $\sim 0.5$ dex relative to field scaling relations, corresponding to gas fractions that are $20- 40\%$ of those found in typical main-sequence galaxies. Although the uncertainties are significant, this suggests that environmental processes efficiently deplete molecular gas reservoirs in the galaxies of this group. The 30m observations place an upper limit on the molecular gas associated with the extended ionized structure, $M_{\rm gas} < 2 \times 10^{10}$ \msun, implying that less than a third of the gas in the intragroup medium is in a cold, star-forming phase. Together, these results show how environmental mechanisms in dense group environments act to remove or suppress molecular gas within galaxies, capturing quenching processes in action.
   	}


   \maketitle

\section{Introduction}

The role of environment on the growth and mass assembly processes of galaxies remains an open question. In particular, the mechanisms by which dense environments, such as groups and clusters, impact the star formation rate (SFR) and its quenching are not yet fully understood. The regulation of star formation may be driven by internal galactic properties, such as stellar mass and morphology, or by external (i.e., environmental) conditions, such as gas stripping and tidal interactions. 

Observations have shown that star formation in galaxies mostly occurs along the main sequence (MS), an evolving  and relatively tight empirical relation between the SFR and the stellar mass  ($\rm M_\star$) of galaxies \citep{Daddi2007, Noeske2007, Elbaz2007, Rodighiero2010, Whitaker2014, Speagle2014}. Such a robust relation, observed so far up to $z \sim 9$ \citep[e.g.,][]{Merida2026}, promotes an overall smooth and continuous mode for star formation in galaxies.  A long-lived star formation cycle could be sustained by a continuous supply of gas and/or minor mergers \citep{Keres2005, Dekel2006, Ocvirk2008, Dekel2009, Genel2010}, providing large reservoirs of molecular gas to fuel star formation \citep{Daddi2010, Tacconi2010, Tacconi2013, Tacconi2018, Sargent2014, Genzel2015, Freundlich2019}. Typical star-forming galaxies are expected to evolve along the MS in a quasi-equilibrium between inflows, outflows, and star formation \citep{Bouche2010, Dave2012, Lilly2013}, until their star formation is quenched when they reach a typical stellar mass of $\sim 10^{11}$\msun\  \citep{Peng2010}. Then, massive galaxies rapidly cease their star formation and drop below the MS to populate the ``red sequence.''

The physical processes responsible for this star formation quenching are not yet fully understood. Quenching could be due to a combination of gas removal by winds or heating from supernovae and/or active galactic nuclei (AGNs) \citep{DiMatteo2005, Maiolino2012, Diamond-Stanic2012, Cicone2014, ForsterSchreiber2014, Muratov2015, Tollet2019}, a decline in gas accretion if inflows cannot penetrate deep inside massive halos \citep{Larson1980, Balogh2000a, Balogh2000b, Dekel2006, Peng2015, Tacchella2016a, Tacchella2016b, Lapiner2023}, a change in morphology \citep{Martig2009}, disk instabilities due to the presence of bars \citep{Gavazzi2015}, and also {environmental} effects, such as gravitational \citep[e.g.,][]{Moore1996} and hydrodynamical perturbations \citep[e.g.,][]{Gunn1972, Larson1980, Abadi1999, Quilis2000, Boselli2022}. 
Both pristine gas accreted from the cosmic web and gas recycled from galaxies through outflows, tidal interactions, or stripping are expected to result in gas reservoirs outside galaxies. The detection of such reservoirs is elusive, in particular because the different gas phases, of different temperature, are mixed together and distributed on different scales \citep[see][for a review]{Boselli2022}. However, dense galactic environments could be the best targets in which to search for them. 
Indeed, large amounts of molecular gas (on the order of $10^{10}$~\msun\ spread over tens of kiloparsecs) have been observed in the intracluster medium of nearby clusters, although the gas is generally associated with well-identified galaxies \citep{Dasyra2012, Verdugo2015, Jachym2017, Moretti2018a, Moretti2018b}. Observations of high-$z$ protoclusters, such as the Spiderweb at $z\sim 2.2$, reveal similar amounts of molecular gas lying between the galaxies \citep{Emonts2016, Emonts2018}.
These recycled gas reservoirs may sustain ongoing star formation and drive the assembly of the massive galaxies found in present-day cluster cores. Similar phenomena are observed on the smaller scale of galaxy groups. 
In Stephan's Quintet, molecular gas is distributed across shocked ridges, tidal tails, and intergalactic regions, forming highly turbulent, unbound giant molecular complexes \citep{Lisenfeld2002, Guillard2012, Maeda2025}.
Given that some of the environmental processes responsible for quenching star formation operate over several gigayears \citep{Cibinel2013, Wetzel2013}, the recently quenched groups observed in the local Universe \citep{Tempel2014, Sohn2016} likely initiated this transition as early as $z > 0.5$.
Furthermore, galaxy groups host the majority of the local galaxy population \citep[e.g.,][]{Eke2004}, making them the most common environment for galactic evolution.

To date, molecular gas measurements in galaxies within dense environments have primarily focused on clusters, brightest cluster galaxies (BCGs), and brightest group galaxies (BGGs). 
In local galaxy clusters, observations have generally found lower molecular gas fractions than in field galaxies and toward denser regions \citep{Fumagalli2009, Boselli2014, Zabel2019, Zabel2022, Morokuma-Matsui2021, Villanueva2022}, although higher gas fractions have also been reported \citep{Nakanishi2006, Mok2016}. At $z\sim 0.7$, \cite{Betti2019} found hints of elevated gas fractions at intermediate densities and lower values at high densities compared to field galaxies using stacked dust continuum measurements. At higher redshifts molecular gas measurements in clusters and protoclusters are more heterogeneous. Numerous studies report gas fractions similar to those of field galaxies \citep{Lee2017,Wang2018, Zavala2019, Williams2022, Gururajan2025}, although often with a relatively large scatter, while other studies report lower \citep{Coogan2018, Alberts2022} or higher \citep{Hayashi2018, Tadaki2019, Gomez-Guijarro2019} values. This heterogeneity could stem from different evolutionary stages, but also from differences in the gas tracers and sample selection. 
While observations demonstrate that central cluster galaxies can host substantial molecular gas reservoirs, with masses reaching $\sim$$10^{10}\rm M_\odot$ or even $\sim$$10^{11}\rm M_\odot$ \citep{Edge2001, Salome2003, McNamara2014, Russell2014, McDonald2014, Fogarty2019, Dunne2021}, these galaxies are often fueled by filaments of accreting cold gas, making them unrepresentative of the broader cluster population. For example, \cite{Castignani2025} mapped the CO molecular gas around three BCGs at $z\sim 0.4$ and found evidence for accreting filaments together with tails of molecular gas due to tidal interactions and stripping.
Comparing their samples with scaling relations, \citet{Castignani2020a, Castignani2020c} inferred lower molecular gas fractions in the majority of their targeted BCGs than in typical star-forming galaxies, possibly due to enhanced star formation followed by gas depletion. However,  other BCG studies infer molecular gas fractions in line with those of typical star-forming galaxies or even higher \citep{Noble2017, Castignani2019, Castignani2022, Castignani2023}, highlighting the diversity in BCG molecular gas properties. 
While several studies reported that group galaxies contain less atomic gas than their field counterparts \citep[e.g.,][]{Catinella2013, Denes2016, For2021, Kleiner2021, Roychowdhury2022}, molecular gas in these environments remains relatively underexplored, limiting our understanding of how the group environment influences this more compact component. In a study of H$_2$-bright galaxies within compact groups, \cite{Alatalo2015} find evidence of star formation suppression despite the presence of substantial molecular gas reservoirs. This suggests that group galaxies may not necessarily need to expel their molecular gas to quench; instead, shocks within the group environment could inject turbulence and thus inhibit star formation. In contrast, \cite{Lee2022} find reduced star formation rates and low molecular gas fractions in two nearby groups. 
More recently, \citet{Toni2026} investigated molecular gas reservoirs in three BGGs at $z\sim0.3$ and interpret their observations (one detection and two upper limits) as an indication of molecular gas depletion in these galaxies. 

\begin{figure}
	\centering
	\includegraphics[width=1\linewidth,trim={0cm 0.cm 0cm 0.cm},clip]{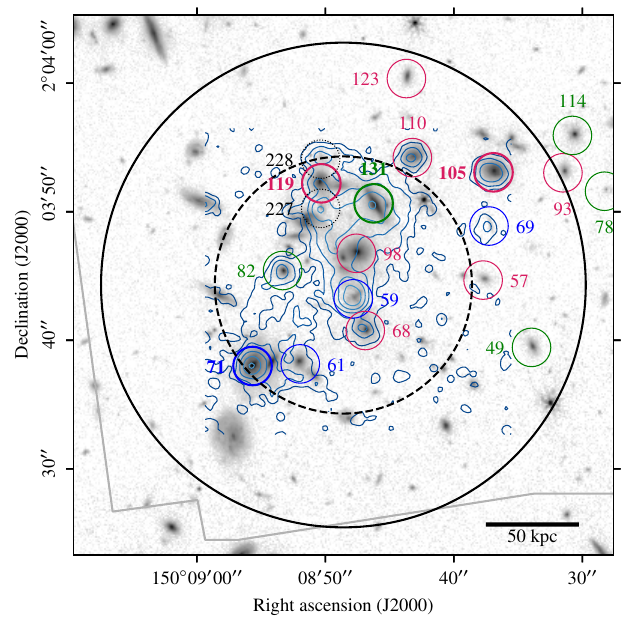}
	 \caption{
    Image obtained with JWST (F277W band, logarithmic scale, arbitrary units) of the over-dense region in the CGr30 galaxy group at $z\sim 0.7$, from the COSMOS-Web program \citep{Casey2023}. Galaxy group members are marked by circles. Logarithmic MUSE \oii\ contours from \citet{Epinat2018, Epinat2024} are overlaid in blue. Eight galaxies in the group, including an AGN host (\#71), are embedded in a huge \oii -bright structure of ionized gas extending over $\sim 10^4$ kpc$^2$. 
	Both the NOEMA primary beam (solid black circle of 38\arcsec\ diameter at $\rm \nu_{obs} = 133.6~GHz$) and the 30m beam (dashed black circle of 20\arcsec\ diameter at the same observing frequency) cover these galaxies. 
    Galaxies observed by MUSE are identified by their MUSE IDs and colored according to their position with respect to the MS (on the MS in green, below in red, and above in blue; cf. Fig.~\ref{fig:MS}). Galaxies whose CO(2-1) emission was detected as part of our NOEMA program, namely \#71, \#105, \#119, and \#131, are highlighted in bold. The MUSE field of view is indicated by a light gray line. 
	}
	\label{fig:targetsfield}
\end{figure}

In this article, we report molecular gas observations of a remarkable galaxy group at $z\sim 0.7$ with the Institut de Radioastronomie Millimetrique (IRAM) NOrthern Extended Millimeter Array (NOEMA) and 30m telescope.  This study aims to characterize the regulation of star formation and quenching processes in a dense environment at intermediate redshift, between the peak epoch of star formation and today. Specifically, we quantified the molecular gas reservoirs within the group galaxies, assessed their depletion relative to those of typical MS galaxies, and investigated the potential presence of diffuse molecular gas within the intragroup medium.
We selected the target group, COSMOS-Gr30 (or CGr30) from the Muse gAlaxy Groups In COSMOS (MAGIC) survey \citep{Epinat2024} because its most overdense region  contains a massive and extended structure of \oiid-bright ionized gas \citep{Epinat2018}, as shown in Fig.~\ref{fig:targetsfield}. 
The present NOEMA and 30m observations provide insight into the overall molecular gas content, both in the group galaxies and in the intragroup medium, which could potentially be available for future star formation. 
Section~\ref{section:obs} presents the target galaxy group and the observations. Section~\ref{section:results} presents the results. Section~\ref{section:conclusion} discusses the results and concludes. Appendix~\ref{appendix:cw} further discusses the position of the group within the cosmic web; appendix~\ref{appendix:SFR_OII} considers alternative \oii\ SFR estimates and appendix~\ref{appendix:sfh} presents the star formation history (SFH) of the group galaxies, where available.
Throughout this paper, we assume a flat $\Lambda$CDM universe with $\Omega_m = 0.3$, $\Omega_\Lambda = 0.7$, and $H_0 =70 \rm~ km s^{-1} Mpc^{-1}$. 

\section{Observations}
\label{section:obs}

\subsection{COSMOS-Gr30}

We selected  COSMOS-Gr30 (or CGr30, 10:00:35.2 +02:03:44.0 J2000), from a sample of 14 groups at intermediate redshift ($0.3<z<0.8$) within the Cosmic Evolution Survey (COSMOS) field \citep{Scoville2007}. This sample was drawn from the zCOSMOS 20k group catalog \citep{Knobel2012} and observed  with the  Multi Unit Spectroscopic Explorer \citep[MUSE; ][]{Bacon2015} as part of the MAGIC survey \citep[][]{Epinat2024}. This survey was designed to explore the impact of environment on galaxy evolution over the last 8 billion years, covering a wide range of environments from relative isolation to dense groups and clusters. 
The CGr30 group shows the highest local overdensity in the sample, with 44 galaxies\footnote{Throughout the paper, we use the IDs from the MAGIC public catalog \citep{Epinat2024}. In this catalog, the IDs are provided in the format X\_CGrY, where X is the galaxy ID in the MUSE fields targeting the COSMOS group with ID Y in the zCOSMOS 20k group catalog \citep{Knobel2012}. Since we only targeted CGr30, we omitted the group name in those IDs. The MAGIC catalog also provides the ID in the COSMOS2015 catalog \citep{Laigle2016} when available.} unambiguously identified spectroscopically with redshifts between $0.719 \leq z \leq 0.732$ in the 1 arcmin$^2$ MUSE field of view (FoV) and a velocity dispersion of $\sim 400$~\kms . This relatively high group velocity dispersion makes this group particularly suited to studying environmental quenching since its galaxies are expected to experience environmental processes over a few gigayears before merging. 
Most importantly, deep MUSE observations reveal a large and massive structure of \oiid-bright ionized gas, extending over $\sim 10^4$ kpc$^2$ and weighting $\sim 5 \pm 3\times 10^{10}$ \msun\ \citep{Epinat2018}.  
Eight of the group galaxies (approximately one-fifth of the members) are part of this overdensity, are embedded in the gaseous structure (see Fig.~\ref{fig:targetsfield}), and seem to constitute a subgroup with a velocity dispersion of about 200~\kms . 
The \oii-emitting diffuse gas is not uniform and displays some filaments and more concentrated regions, although these are not necessarily centered on the galaxies involved in the structure. 
This suggests that it was extracted from galaxies through tidal forces, outflows due to star formation and AGN activity, and possibly ram-pressure stripping due to previously-stripped material or more pristine gas from the cosmic web. Indeed, CGr30 is located in a larger-scale structure identified as the COSMOS-Wall \citep{Iovino2016}, at the intersection of cosmic web filaments, as shown in Appendix~\ref{appendix:cw}.
Nevertheless, the group galaxies do not show clear morphological signs of interactions \citep[cf. Fig.~\ref{fig:targetsfield} and][]{Epinat2018}.
As described in \cite{Epinat2018}, \oii\ kinematics separate two main regions in the extended gas structure (north-east versus south-west in Fig.~\ref{fig:targetsfield}). The velocity offset between the two is larger than $\sim 200\rm ~km/s$, and each region seems to be kinematically linked with a massive galaxy, namely \#98 and \#71 ($\rm M_{\rm star}\approx 10^{10.5-11}$\msun). Among these two galaxies, we note that \#71 is an AGN host, as shown by broad emission lines in the nucleus, a bright [\ion{O}{iii}] line with respect to \oii\ and \hbeta, [\ion{Ne}{v}] and [\ion{Ne}{iii}] lines, and X-ray emission \citep[][section 5.1.3]{Epinat2018}.
The other galaxies are star-forming and less massive ($\rm M_{\rm star} \approx 10^{9-10.5}$\msun).

Since the surveyed group lies in the COSMOS field, it benefits from ancillary data that enable us to infer different galaxy physical properties. 
We used the stellar masses, SFRs, and extinctions derived in \citet{Epinat2024} from spectral energy distribution (SED) models. These models were constrained with the Code Investigating GALaxy Emission \citep[\textsc{Cigale}; ][]{Boquien2019} using the MUSE spectroscopic redshifts and the COSMOS2020 photometric measurements in 3\arcsec\ diameter apertures in 32 bands, from UV to radio centimeter  \citep{Weaver2022}. 
As explained in more details in \cite{Epinat2024}, the analysis used single stellar populations from \citet{Bruzual2003} with a \citet{Salpeter1955} initial mass function and a fixed metallicity of 0.02 dex. The SFH is a delayed exponential law with flexibility in the most recent period, as described in \citet{Ciesla2017, Ciesla2021}. The attenuation law is a modified \citet{Calzetti2000} law with a fixed total-to-selective extinction ratio $R_V = 3.1$, and the dust template model is that of \citet{Dale2014}. 
We compared the resulting stellar masses with values obtained using \textsc{Fast} \citep{Kriek2009} and found good agreement. For the  \textsc{Fast} analysis, we used a synthetic library generated from the stellar population synthesis models of \cite{Conroy2010}, a \citet{Chabrier2003} initial mass function, an exponentially declining SFR, and a \citet{Calzetti2000} extinction law, as in \citet{Mercier2022}. 
As is common for SED-based SFR estimates \citep[e.g.,][]{Wuyts2011a, Leja2018}, the values from \citet{Epinat2024} involve significant uncertainties, which we account for in our analysis. 
To ensure consistency between the MUSE and NOEMA data, we also convolved the MUSE datacube to the spatial resolution of NOEMA, assuming an elliptical Gaussian beam, and extracted both NOEMA and MUSE spectra at the same sky positions.
We used \texttt{pPXF} \citep{Cappellari2004, Cappellari2017,Cappellari2023} to determine the line-of-sight velocity of the stars using the E-MILES stellar population synthesis templates \citep{Vazdekis2016} and to remove the continuum contribution from the spectra, following the same strategy as in \cite{Epinat2018}. 
We used multiplicative polynomials of order ten to fit the continuum. Such polynomials are well suited to removing the continuum with a reduced impact on the shape of emission lines. We checked that they provide similar line-of-sight velocities to those obtained using additive polynomials, within less than $\pm14$~\kms . We also note that, although the spectral resolution of the E-MILES templates is larger than that of MUSE at the redshift of CGr30, this did not prevent us from obtaining accurate fits as long as the observed velocity dispersion of the integrated spectra remained larger than the template resolution, which corresponds to about $\sim 50$~\kms. This is the case in practice for all galaxies except \#57, \#93, and \#123, although a visual inspection of the continuum-subtracted spectra suggests that the fits are correct.
We then fitted the \oii\ doublet for each MUSE continuum-subtracted spectrum using the 1D function of the \texttt{Camel} code\footnote{\url{https://gitlab.lam.fr/bepinat/CAMEL}} to infer velocities with respect to the systemic redshift of the $z=0.7251$ group, as well as the doublet velocity dispersion.
To infer the offset from the MS, we adopted the ${\rm SFR}(MS;z,M_\star)$ parametrization derived by \citet{Mercier2022} for the MAGIC sample, applying a downward adjustment of 0.12 dex to account for the average offset between their \oii\ SFRs and the SED SFRs of \citet{Epinat2024}\footnote{The \oii\ SFRs of \citet{Mercier2022} and the SED SFRs of \citet{Epinat2024} were both measured within the same 3\arcsec\ apertures, but the \oii\ SFRs may be unreliable for CGr30 galaxies lying in the \oii-emitting diffuse structure because of possible contamination from the structure. Hence, we used the SED SFRs of \citet{Epinat2024} despite their large uncertainties. In Appendix~\ref{appendix:SFR_OII}, we also consider \oii\ SFRs.}.
Table~\ref{table:properties} lists some of the properties of the 17 CGr30 galaxies falling within the NOEMA primary beam. Four of these galaxies (\#57, \#61, \#93, and \#123) were not detected in \oii\ with MUSE, while the two lowest-mass galaxies (\#69, \#78) do not yield secure \texttt{pPXF} fits.

\begin{table*}
		\caption{
		Properties of the 17 galaxies of COSMOS-Gr30 within the NOEMA primary beam. 
	}      
	\label{table:properties}      
	\centering          
	\footnotesize
\begin{tabular*}{\textwidth}{@{\extracolsep{\fill}} rrrrrrrrrrr}  
    \hline\hline \\[-0.3cm]
    \multicolumn{1}{l}{${\rm ID}$}  & 
    \multicolumn{1}{l}{RA} & 
    \multicolumn{1}{l}{DEC} & 
    \multicolumn{1}{l}{$z$} & 
    \multicolumn{1}{l}{$\log M_\star$} & 
    \multicolumn{1}{l}{$\log {\rm SFR}$} & 
    \multicolumn{1}{l}{$\varv_{\rm pPXF} $} & 
    \multicolumn{1}{l}{$\varv_{\rm \oii} $} & 
    \multicolumn{1}{l}{${\rm FWHM_{\rm \oii}}$} & 
    \multicolumn{1}{l}{$\delta {\rm MS}$} & 
    \multicolumn{1}{l}{\!\!$\log M_{\rm exp}$}\\
    
    & & & & 
    \multicolumn{1}{l}{$\rm [M_\odot]$} & 
    \multicolumn{1}{l}{$\rm [M_\odot {\rm yr}^{-1}]$} & 
    \multicolumn{1}{l}{$[{\rm km\ s^{-1}}]$} & 
    \multicolumn{1}{l}{$[{\rm km\ s^{-1}}]$} & 
    \multicolumn{1}{l}{$[{\rm km\ s^{-1}}]$} & 
    & 
    \multicolumn{1}{l}{\!\!$\rm [M_\odot]$}\\
    \hline \\[-0.3cm]      
    49  & 10:00:34.2648 & +02:03:39.456 & $0.7244$ & $9.42 \pm 0.06$  & $-0.02\pm 0.86$ & $-24$  & $-123$ & $124$ & $0.14$  & $9.51$  \\
    57  & 10:00:34.5144 & +02:03:44.676 & $0.7251$ & $9.18 \pm 0.03$  & $-2.90\pm 2.01$ & $-126$ &        &       & $-2.56$ & $7.99$  \\
    59  & 10:00:35.1888 & +02:03:43.236 & $0.7231$ & $9.19 \pm 0.06$  & $0.97\pm 0.32$  & $-316$ & $-335$ & $163$ & $1.31$  & $9.96$  \\
    61  & 10:00:35.4648 & +02:03:38.160 & $0.7239$ & $10.06 \pm 0.11$ & $1.09\pm 0.27$  & $-196$ &        &       & $0.79$  & $10.22$ \\
    68  & 10:00:35.1264 & +02:03:40.752 & $0.7260$ & $10.28 \pm 0.08$ & $-0.37\pm 2.67$ & $124$  & $-283$ & $279$ & $-0.83$ & $9.52$  \\
    69  & 10:00:34.4856 & +02:03:48.888 & $0.7255$ & $8.30 \pm 0.19$  & $-0.30\pm 1.91$ &        & $82$   & $79$  & $0.67$  & $9.12$  \\
    71  & 10:00:35.7120 & +02:03:38.016 & $0.7249$ & $10.94 \pm 0.05$ & $2.15\pm 0.47$  & $-46$  & $-34$  & $436$ & $1.22$  & $10.95$ \\
    78  & 10:00:33.8856 & +02:03:51.588 & $0.7271$ & $8.12 \pm 0.11$  & $-1.38\pm 2.03$ &        & $302$  & $92$  & $-0.29$ & $8.53$  \\
    82  & 10:00:35.5560 & +02:03:45.432 & $0.7268$ & $10.10 \pm 0.04$ & $0.42\pm 0.46$  & $279$  & $279$  & $393$ & $0.09$  & $9.89$  \\
    93  & 10:00:34.1016 & +02:03:53.028 & $0.7250$ & $9.70 \pm 0.02$  & $-3.42\pm 4.11$ & $-105$ &        &       & $-3.46$ & $7.84$  \\
    98  & 10:00:35.1720 & +02:03:46.764 & $0.7250$ & $11.09 \pm 0.08$ & $0.14\pm 0.28$  & $-115$ & $-42$  & $535$ & $-0.90$ & $9.96$  \\
    105 & 10:00:34.4616 & +02:03:53.064 & $0.7269$ & $10.46 \pm 0.06$ & $-0.37\pm 1.28$ & $319$  & $319$  & $280$ & $-0.95$ & $9.56$  \\
    110 & 10:00:34.8816 & +02:03:54.216 & $0.7267$ & $10.06 \pm 0.10$ & $-0.24\pm 0.75$ & $317$  & $309$  & $97$  & $-0.54$ & $9.54$  \\
    114 & 10:00:34.0512 & +02:03:55.944 & $0.7242$ & $9.61 \pm 0.17$  & $-0.09\pm 0.81$ & $-106$ & $-110$ & $171$ & $-0.06$ & $9.52$  \\
    119 & 10:00:35.3544 & +02:03:52.200 & $0.7247$ & $10.60 \pm 0.07$ & $0.06\pm 2.02$  & $-153$ & $-173$ & $141$ & $-0.63$ & $9.81$  \\
    123 & 10:00:34.9128 & +02:04:00.372 & $0.7234$ & $9.74 \pm 0.07$  & $-2.00\pm 1.59$ & $-393$ &        &       & $-2.07$ & $8.57$  \\
    131 & 10:00:35.0832 & +02:03:50.652 & $0.7230$ & $10.73 \pm 0.12$ & $0.66\pm 0.76$  & $-405$ & $-253$ & $622$ & $-0.12$ & $10.15$ \\
    \hline                  
\end{tabular*}
\tablefoot{ID numbers correspond to those of CGr30 galaxies from the MAGIC public catalog \citep{Epinat2024}; \citet{Epinat2024} derived the stellar mass and SFR from SED models; we obtained the stellar velocity with respect to the systemic group redshift $z=0.7251$ using \texttt{pPXF}; \oii\ we obtained the velocity and FWHM from MUSE observations using the 1D function of the \texttt{Camel} code; $\delta {\rm MS}={\rm SFR}/{\rm SFR}({\rm MS}; z, M_\star)$ gives the offset from the MS, which we calculated using the \citet{Mercier2022} parametrization of the MS with a 0.12 dex downward adjustment to account for the offset between the \oii\ SFRs of \citet{Mercier2022} and the SED-based estimates of \citet{Epinat2024}; we calculated the expected molecular gas mass, $M_{\rm exp}$, using the \citet{Tacconi2020} parametrization of the molecular gas fraction, $\mu_{\rm gas}(z,M_\star,\delta{\rm MS})$ (cf. Eq.~\ref{eq:Tacconi}). 
Four galaxies (\#57, \#61, \#93, \#123) were not detected in \oii\ with MUSE; the two lowest-mass galaxies (\#69, \#78) do not yield secure \texttt{pPXF} fits.}
\end{table*}

\begin{figure}
	\centering
	\includegraphics[width=1\linewidth,trim={0.5cm 0.2cm 1cm 1.5cm},clip]{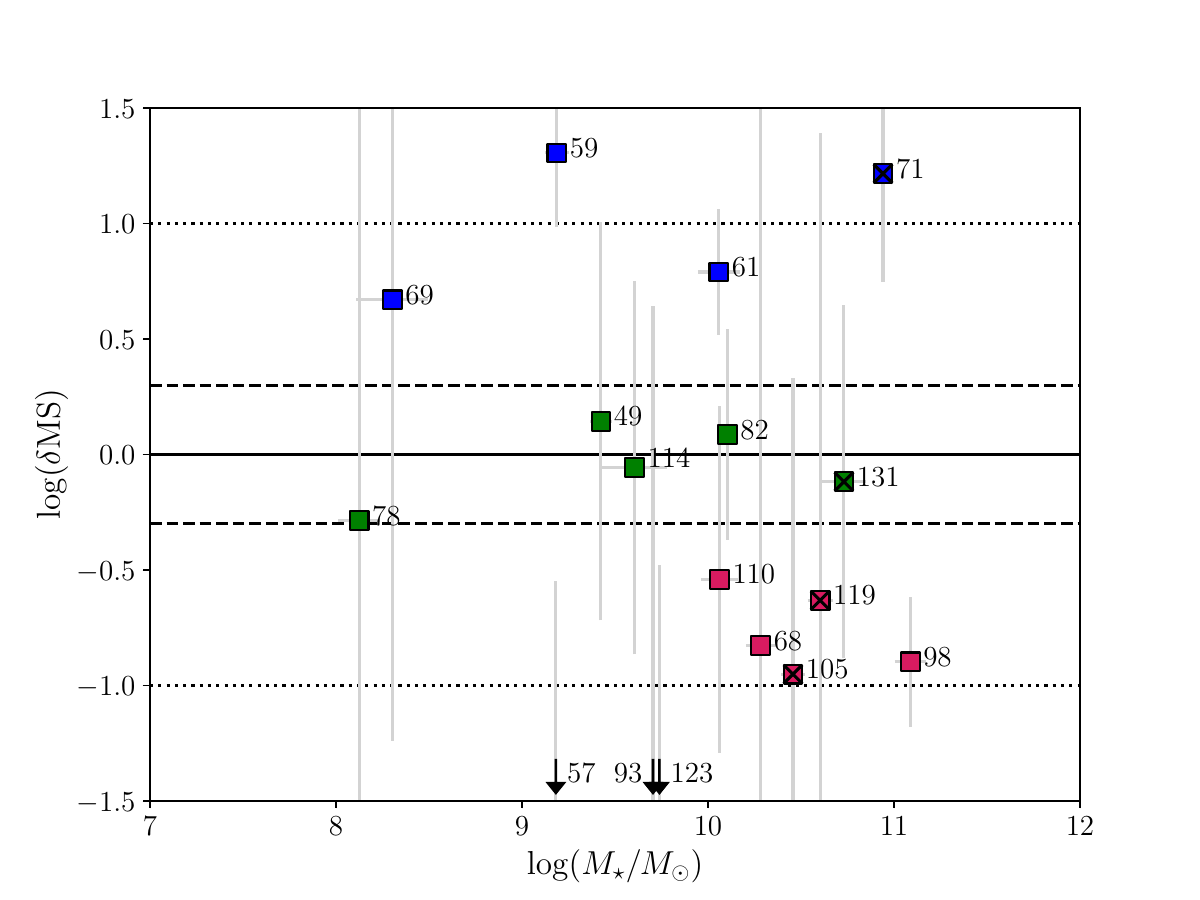}
	\caption{
		Position of the 17 galaxies of COSMOS-Gr30 galaxies within the NOEMA primary beam with respect to the MS. The offset from the MS ($\delta {\rm MS} = {\rm SFR/SFR}({\rm MS};z,M_\star)$) is plotted as a function of stellar mass ($M_\star$), using the reference ${\rm SFR}({\rm MS};z,M_\star)$ from \citet{Mercier2022} with a downward adjustment of 0.12 dex. The solid, dashed, and dotted horizontal lines represent the MS ridge, its 0.3 dex scatter, and $\pm 1$ dex from the MS, respectively. Galaxies with detected CO(2-1) emission are indicated by crosses. Error bars represent the uncertainties in $M_\star$ and SFR. 
	}
	\label{fig:MS}
\end{figure}

\subsection{NOEMA observations}

To probe the molecular gas content of CGr30 galaxies at $z\sim0.725$, we targeted the $^{12}$CO(2-1) carbon monoxide emission line ($\rm \nu_{rest} = 230.538~GHz$) in NOEMA's 2mm band. This line allows direct comparison with other molecular gas surveys, such as the Plateau de Bure high-z Blue Sequence Survey \citep[PHIBSS2; ][]{Freundlich2019} and is the lowest-J CO line available with NOEMA at the galaxies’ redshifts. Since the targets are all separated by at least 2\arcsec , the observations were carried out in configuration C to achieve a spatial resolution of $\sim$1.9\arcsec, which does not resolve individual galaxies. The NOEMA primary beam of 38\arcsec\ encompasses 17 group members with secure redshifts, stellar masses, and SFRs. Table~\ref{table:properties} lists some of their properties, and Fig.~\ref{fig:MS} shows their location with respect to the MS in the stellar mass--SFR plane.
The observations took place at the NOEMA interferometer at the Plateau de Bure in France between January and March 2020, using ten antennas, as part of the observational program S19BV (PI: Contini, contact: Freundlich), for a total on-source observing time of 12.2 hours. 
The weather conditions varied, with system temperatures ranging between 100 and 300 K depending on atmospheric conditions, wind speeds below $20 ~\rm m\ s^{-1}$, and precipitable water vapor between 2 and 6 mm. The absolute flux scale was derived from secondary flux calibrators (3C84, LKHA101, J1028-0236, 1038+064, 3C279, 1055+018, MWC349, 0851+202, 0906+015, and J0948+003), whose fluxes are regularly measured using Jupiter's satellites or planets. We calibrated the data using the \texttt{clic} software of the IRAM \texttt{gildas} package and further analyzed and mapped using the \texttt{gildas mapping} and \texttt{class} software.
The resulting beam size is $2.6\times1.1~\rm arcsec^2$ and the root mean square (RMS) noise is $\sigma_{\rm RMS}=0.26~\rm mJy$ over velocity channels of width $\Delta v = 44.86~\rm km s^{-1}$, obtained using \texttt{gildas mapping go noise} tool. 
 
\begin{table*}
	\caption{
        NOEMA CO(2-1) measurements
	}             
	\label{table:CO}      
	\centering          
	\footnotesize
\begin{tabular*}{\textwidth}{@{\extracolsep{\fill}} rrrrrrrrrrr}
    \hline\hline 
    \\[-0.3cm]
    \multicolumn{1}{l}{${\rm ID}$} & 
    \multicolumn{1}{l}{$F({\rm CO})$} & 
    \multicolumn{1}{l}{$dF({\rm CO})$} & 
    \multicolumn{1}{l}{$\rm FWHM_{\rm CO}$} & 
    \multicolumn{1}{l}{$d\rm FWHM_{\rm CO}$} & 
    \multicolumn{1}{l}{$\varv_{\rm CO}$} & 
    \multicolumn{1}{l}{$d \varv_{\rm CO}$} & 
    \multicolumn{1}{l}{$\log L_{\rm CO(2-1)}$} & 
    \multicolumn{1}{l}{$\log M_{\rm gas}$} & 
    \multicolumn{1}{l}{$\log \mu_{\rm gas}$} & 
    \multicolumn{1}{l}{$\log t_{\rm depl}$}\\
    
    & 
    \multicolumn{1}{l}{$[\rm Jy~km~s^{-1}]$} & 
    \multicolumn{1}{l}{$[\rm Jy~km~s^{-1}]$} & 
    \multicolumn{1}{l}{$[\rm km~s^{-1}]$} & 
    \multicolumn{1}{l}{$[\rm km ~s^{-1}]$} & 
    \multicolumn{1}{l}{$[\rm km ~s^{-1}]$} & 
    \multicolumn{1}{l}{$[\rm km ~s^{-1}]$} & 
    \multicolumn{1}{l}{$[\rm K ~km ~s^{-1} ~pc^2]$} & 
    \multicolumn{1}{l}{$\rm [M_\odot]$} & 
    & 
    \multicolumn{1}{l}{[Gyr]}\\
    \hline \\[-0.3cm]                    
    49  & $< 0.06$ &      &     &    &      &    & $<8.61$ & $<9.58$  & $<0.15$  & $<0.60$  \\
    57  & $< 0.06$ &      &     &    &      &    & $<8.60$ & $<9.65$  & $<0.47$  & $<3.55$  \\
    59  & $< 0.07$ &      &     &    &      &    & $<8.67$ & $<9.72$  & $<0.53$  & $<-0.25$ \\
    61  & $< 0.06$ &      &     &    &      &    & $<8.60$ & $<9.41$  & $<-0.65$ & $<-0.68$ \\
    68  & $< 0.09$ &      &     &    &      &    & $<8.79$ & $<9.56$  & $<-0.73$ & $<0.93$  \\
    69  & $< 0.05$ &      &     &    &      &    & $<8.51$ & $<10.08$ & $<1.78$  & $<1.38$  \\
    71  & $0.43$   & $0.05$ & 566 & 73 & $-46$ & 36 & $9.47$  & $10.17$  & $-0.77$  & $-0.97$  \\
    78  & $< 0.05$ &      &     &    &      &    & $<8.55$ & $<10.29$ & $<2.16$  & $<2.67$  \\
    82  & $< 0.10$ &      &     &    &      &    & $<8.86$ & $<9.66$  & $<-0.44$ & $<0.25$  \\
    93  & $< 0.06$ &      &     &    &      &    & $<8.60$ & $<9.49$  & $<-0.22$ & $<3.91$  \\
    98  & $< 0.12$ &      &     &    &      &    & $<8.93$ & $<9.62$  & $<-1.47$ & $<0.48$  \\
    105 & $0.21$   & $0.04$ & 233 & 46 & 312  & 21 & $9.18$  & $9.92$   & $-0.53$  & $1.29$   \\
    110 & $< 0.05$ &      &     &    &      &    & $<8.56$ & $<9.37$  & $<-0.70$ & $<0.61$  \\
    114 & $< 0.07$ &      &     &    &      &    & $<8.68$ & $<9.59$  & $<-0.01$ & $<0.68$  \\
    119 & $0.19$   & $0.03$ & 173 & 34 & $-140$ & 18 & $9.13$  & $9.86$   & $-0.74$  & $0.80$   \\
    123 & $< 0.06$ &      &     &    &      &    & $<8.60$ & $<9.47$  & $<-0.26$ & $<2.48$  \\
    131 & $0.48$   & $0.06$ & 579 & 79 & $-412$ & 39 & $9.53$  & $10.24$  & $-0.49$  & $0.58$   \\
    \hline                  
\end{tabular*}
\tablefoot{The table lists the integrated CO(2-1) line flux ($F_{\rm CO(2-1)}$), the full width at half maximum ($\rm FWHM_{\rm CO}$), the velocity offset ($\varv_{\rm CO}$) with respect to the mean group velocity and their uncertainties, together with the derived intrinsic CO(2-1) luminosity ($L_{\rm CO(2-1)}$), molecular gas mass ($M_{\rm gas}$), gas-to-stellar mass ratio ($\mu_{\rm gas}= M_{\rm gas}/M_\star$), and depletion time ($t_{\rm depl}=M_{\rm gas}/{\rm SFR}$) for the 17 galaxies in the NOEMA primary beam. Inequalities indicate 3$\sigma$ upper limits within the FWHM. In Fig.~\ref{fig:mutdepl_vs_DMS}, the data points correspond to the median values from the Monte Carlo samplings and not directly to $M_{\rm gas}/M_{\rm exp}$ taken from the tables.}
\end{table*}

\subsection{30m observations}

In addition to the NOEMA observations, we observed the CO(2-1) line with the IRAM 30m telescope at Pico Veleta in Spain to constrain the total molecular gas content in both the galaxies and the intragroup medium. Any excess emission compared to the total galaxy flux measured from the NOEMA observations could reveal molecular gas outside the galaxies and thus help test models of star formation regulation in galaxies. 
These observations were first carried out as part of project 110-19, directly complementing the NOEMA project S19BV, between 30 and 31 July 2020, and then as part of the dedicated project 173-20  (PI: Contini, contact: Freundlich) between 7 and 19 April 2021. The two observing campaigns resulted in a total on-source observing time of 9.5 hours. 
Figure~\ref{fig:targetsfield} shows that the 20\arcsec\ beam of the 30m telescope at the targeted wavelength covers the \oii-emitting diffuse gas structure and the eight galaxies embedded in this structure. Other galaxies (\#57, \#69, \#105, and \#110) lie at the edge of the beam. 
We used band E150 of the Eight Mixer Receiver (EMIR) to observe the CO(2-1) line, wobbler-switching, and both the Wideband Line Multiple Autocorrelator (WILMA) and the Fast Fourier Transform Spectrometers (FTS) as backends. 
We carried out the observations under average weather conditions, with system temperatures of $\rm T_{sys}\sim 150~K$. 
We performed data reduction and analysis using the \texttt{class} software of the \texttt{gildas} package.
The resulting RMS noise was $1.7 ~\rm mJy$ over velocity channels of width $89.74~\rm km~s^{-1}$ with WILMA, after converting the antenna temperature into a main beam temperature using a main beam efficiency $B_{\rm eff}=0.62$ and a point source sensitivity of $5\rm ~Jy/K$ at the observing frequency ($\nu_{\rm obs} = 133.6~\rm GHz$)\footnote{
Cf. \url{https://publicwiki.iram.es/Iram30mEfficiencies}, in particular the IRAM internal reports \citet{Kramer1997}, Eq.~(17), and \citet{Kramer2013}. 
}.

\begin{figure*}
	\centering
	\includegraphics[width=0.49\linewidth,trim={.cm 0cm 0.cm 0cm},clip]{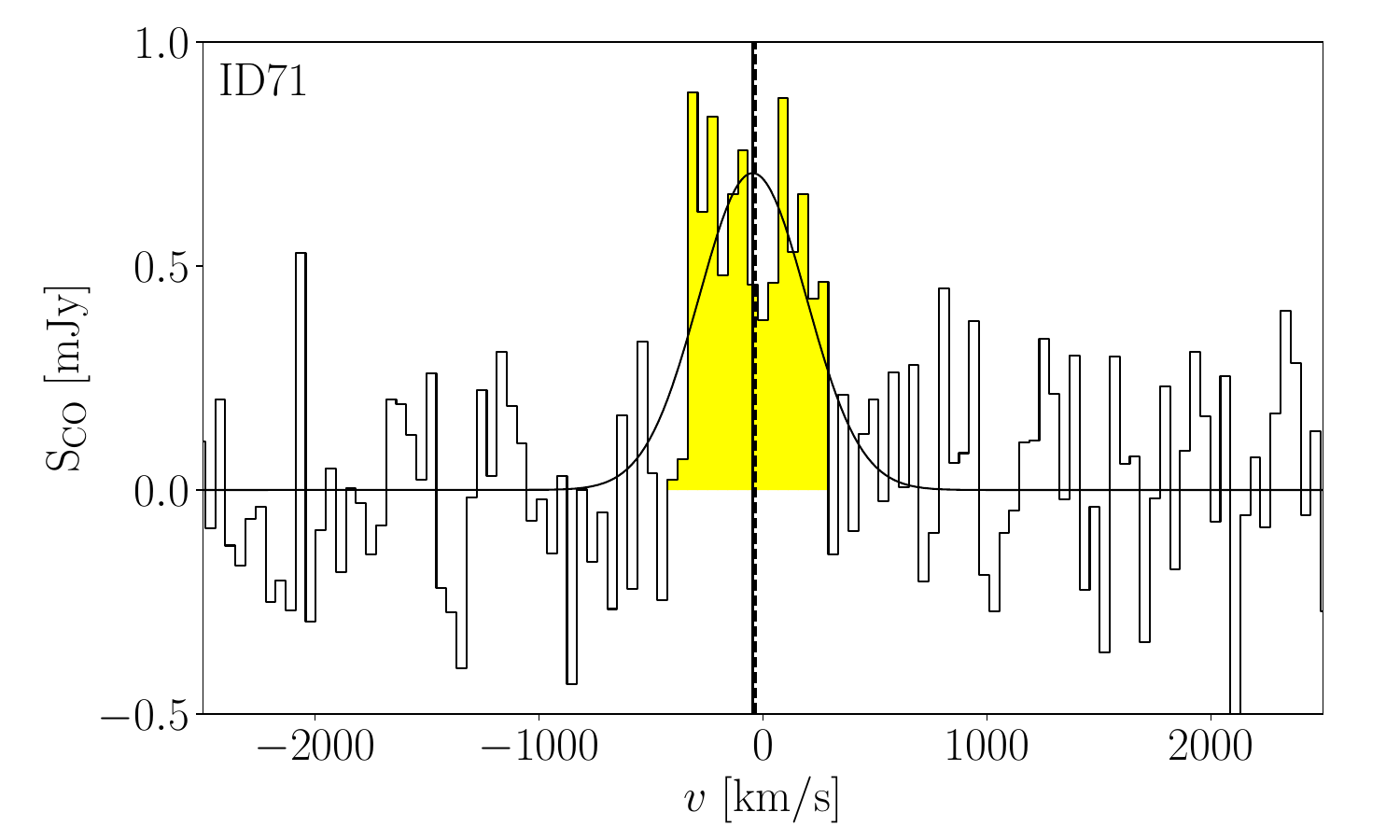}
	\includegraphics[width=0.49\linewidth,trim={.0cm 0cm 0.cm 0cm},clip]{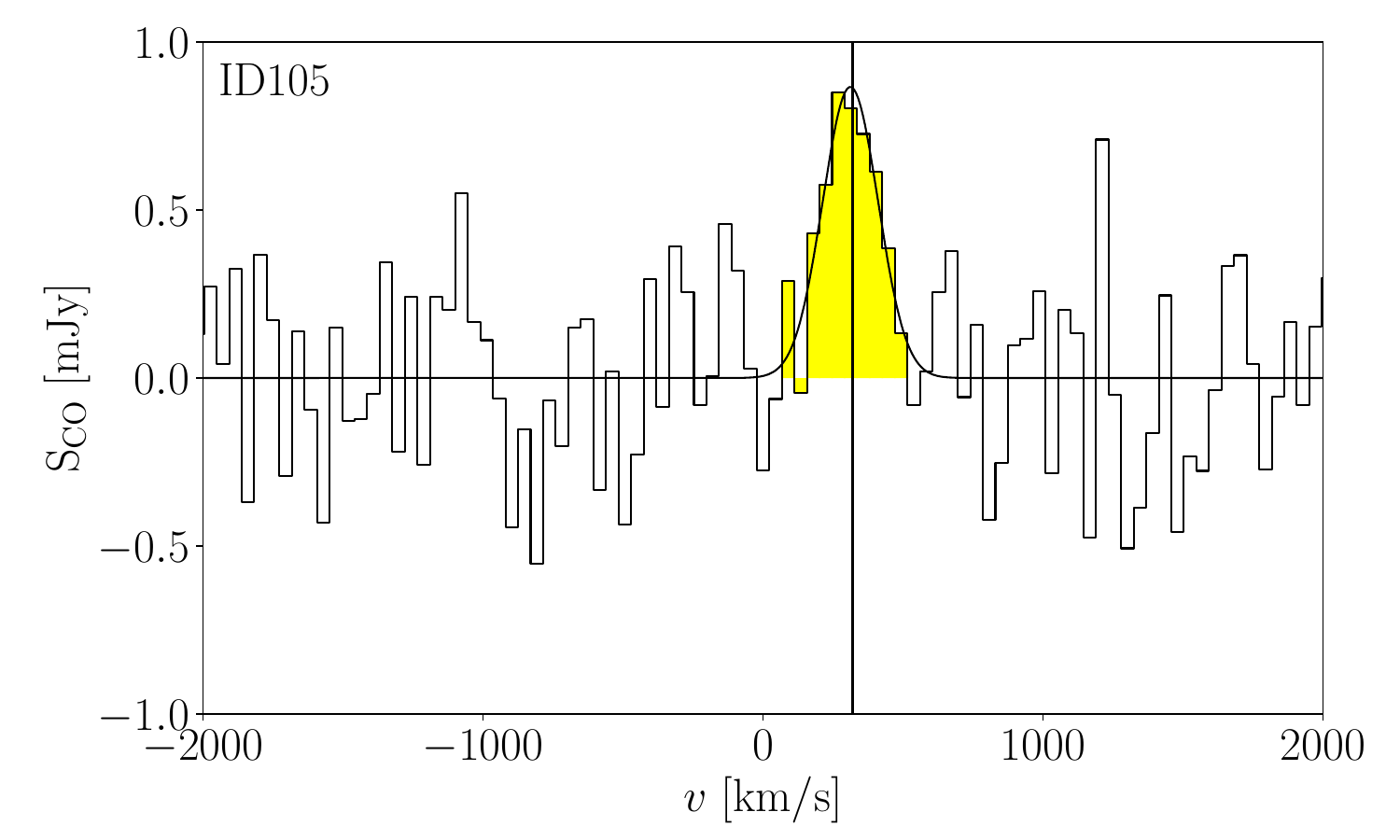}
	
	\includegraphics[width=0.49\linewidth,trim={.0cm 0cm 0cm 0cm},clip]{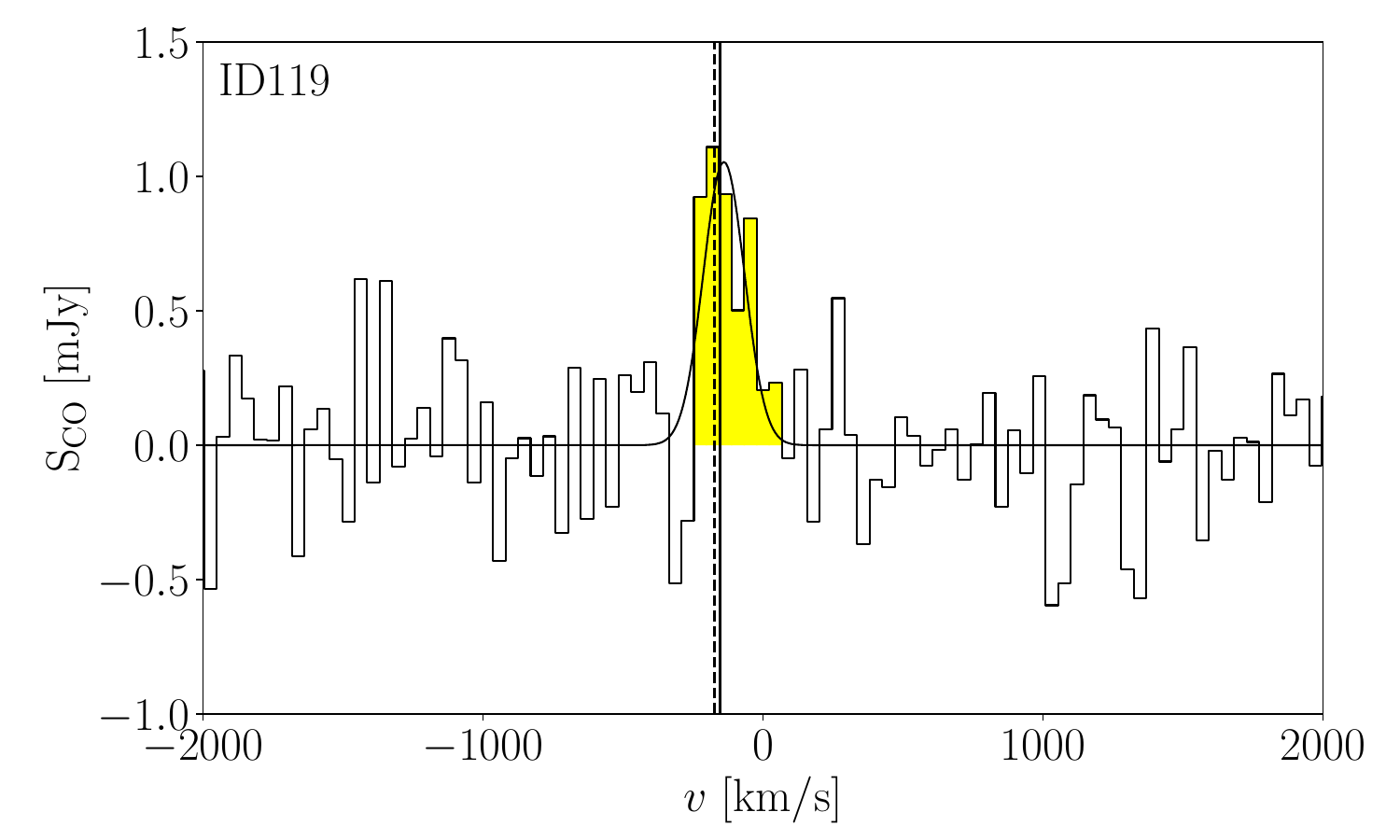}
	\includegraphics[width=0.49\linewidth,trim={.0cm 0cm 0cm 0cm},clip]{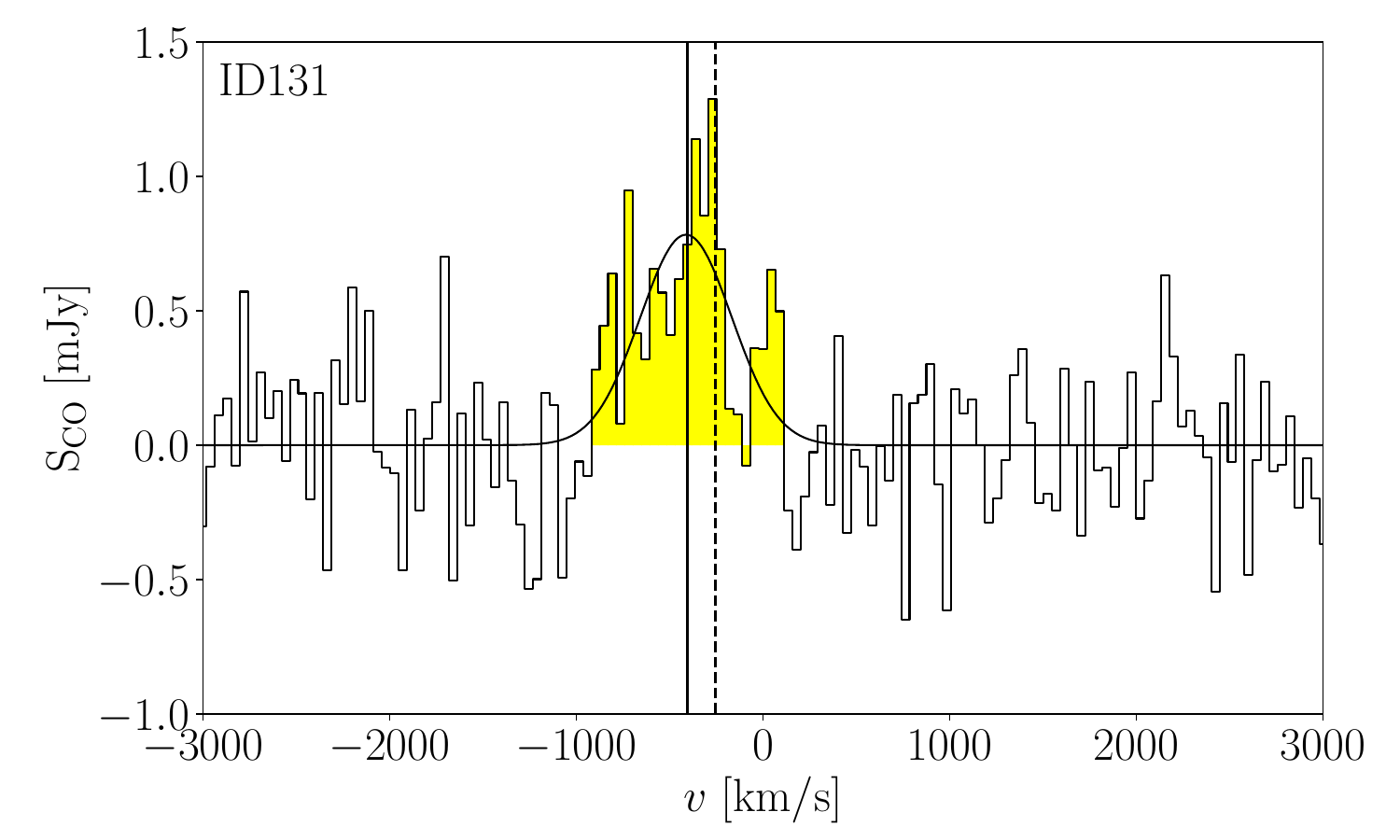}
	\caption{Spectra of the CO(2-1) emission obtained with NOEMA for the four detected galaxies in the CGr30 group, namely \#71, \#105, \#119, and \#131. The S/N values are 8.7, 5.4, 8.6, and 8.3, respectively (cf. Table~\ref{table:CO}). We define the velocities $\varv$ with respect to the average group velocity. We show the stellar velocities as vertical solid lines and those of the \oii\ doublet as vertical dashed lines. 
	}
	\label{fig:detections}
\end{figure*}

\section{Results}
\label{section:results}

\subsection{Detections and upper limits}

For each galaxy of the group, we extracted the NOEMA spectrum at the optical position by fitting the point spread function (PSF) in the $uv$ Fourier space using the  \texttt{gildas mapping go uvfit} tool. Fitting the unresolved sources through their $uv$ visibilities avoids the imaging and deconvolution steps, which require assumptions to compensate for the sparse $uv$ coverage. Table~\ref{table:CO} summarizes the quantities derived from these galaxy spectra. 

We detected the CO(2-1) line with $\rm S/N>5$ for four galaxies, namely \#71, \#105, \#119, and \#131, whose spectra are displayed in Fig.~\ref{fig:detections}. We performed Gaussian fits using \texttt{class}, yielding the corresponding integrated line flux $F({\rm CO})$, velocity $\varv_{\rm CO}$ with respect to the mean group velocity, and full width at half maximum $\rm FWHM_{\rm CO}$. 
We rescaled the fluxes to take into account the NOEMA primary beam power, assuming a Gaussian beam with a 38\arcsec\ half-power beam width.  
For the four detected galaxies, the S/N are 8.7, 5.4, 8.6, and 8.3, respectively. 
Figure~\ref{fig:detections} shows that the velocity of the CO line closely matches that of the stars ($\varv_{\rm pPXF}$ in Table~\ref{table:properties}) and that of the \oii\ emission ($\varv_{\rm \oii}$), although there is a $\sim -160$~\kms\ offset with the latter in the case of \#131. The agreement with $\varv_{\rm pPXF}$ fits well with the idea that the molecular gas lies in the galaxy close to the stars, while the global offset between molecular and ionized gas velocity profiles in the case of \#131 could indicate that some of the \oii\ emission traces ionized gas being decelerated by ram pressure stripping \citep[see e.g.,][]{Boselli2019}. Supporting this, \citet{Epinat2018} report a higher velocity dispersion in the diffuse \oii\ component around \#131, and that the kinematics of this galaxy differ from its surrounding \oii\ gas. The agreement between molecular and ionized gas velocity profile widths for the four galaxies also suggests that the CO and \oii\ emissions have similar extents within these galaxies.
\cite{Munoz-Lopez2025} derived star formation histories for some of the galaxies of CGr30, and we note that the four galaxies we detect harbor a recent star formation episode, as shown in Appendix~\ref{appendix:sfh}, possibly due to an influx of gas triggered by the environment. Indeed, the last star formation episode for these four galaxies took place 1 or 2 Gyr ago, hence about a group dynamical time ($\sim$$1.6\rm ~Gyr$).

\setlength{\unitlength}{\linewidth/100}

\begin{figure}
	\centering
	\begin{picture}(100,60)
		\put(-4,0){ 
			\includegraphics[width=\linewidth,trim={0.cm 0cm 0cm 0cm},clip]{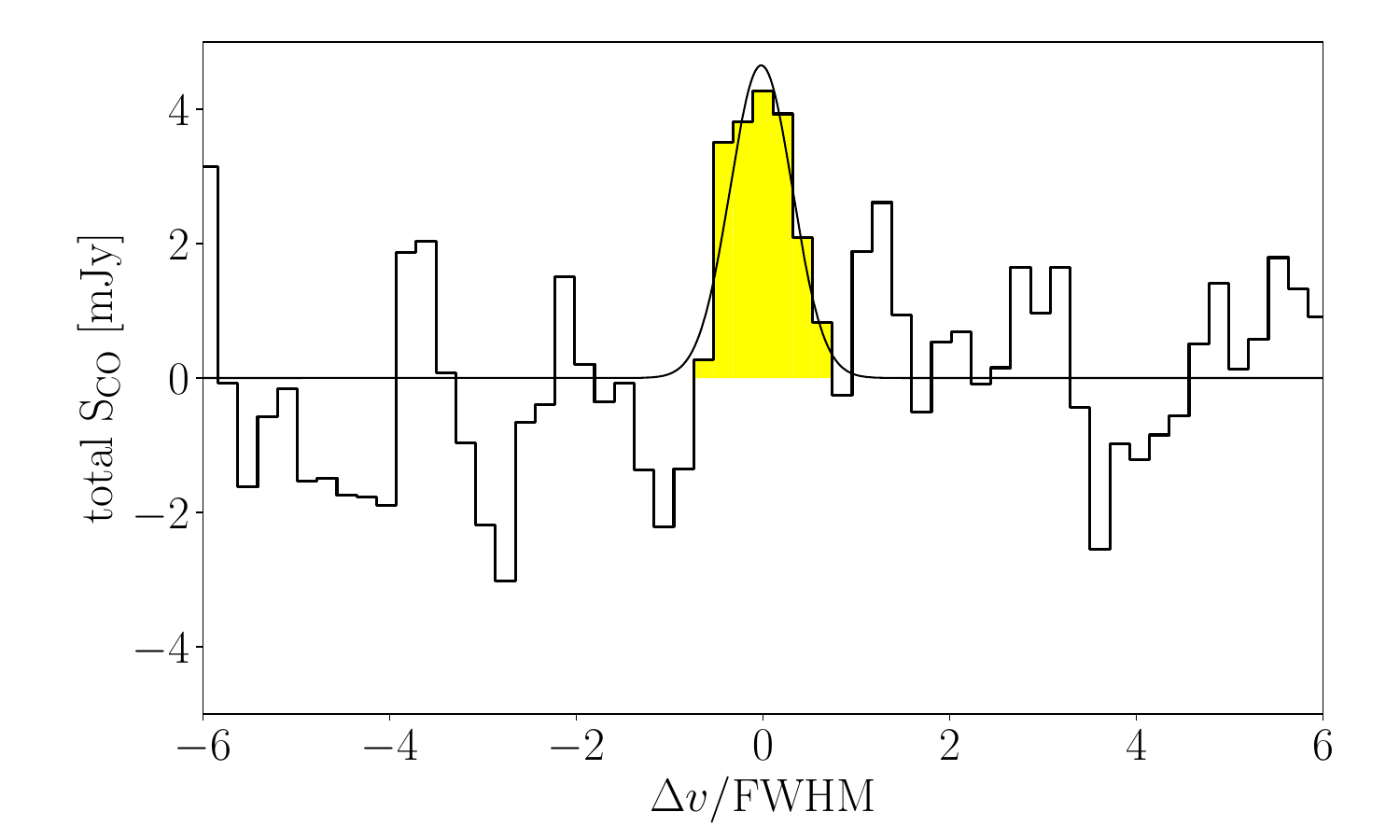}
			}
		\put(14,53){\tiny \scalebox{1.2}{all}}
	\end{picture}
	\\
    \vspace{0.4cm}
    \begin{picture}(100,60)
		\put(-4,0){ 
			\includegraphics[width=\linewidth,trim={0.cm 0cm 0cm 0cm},clip]{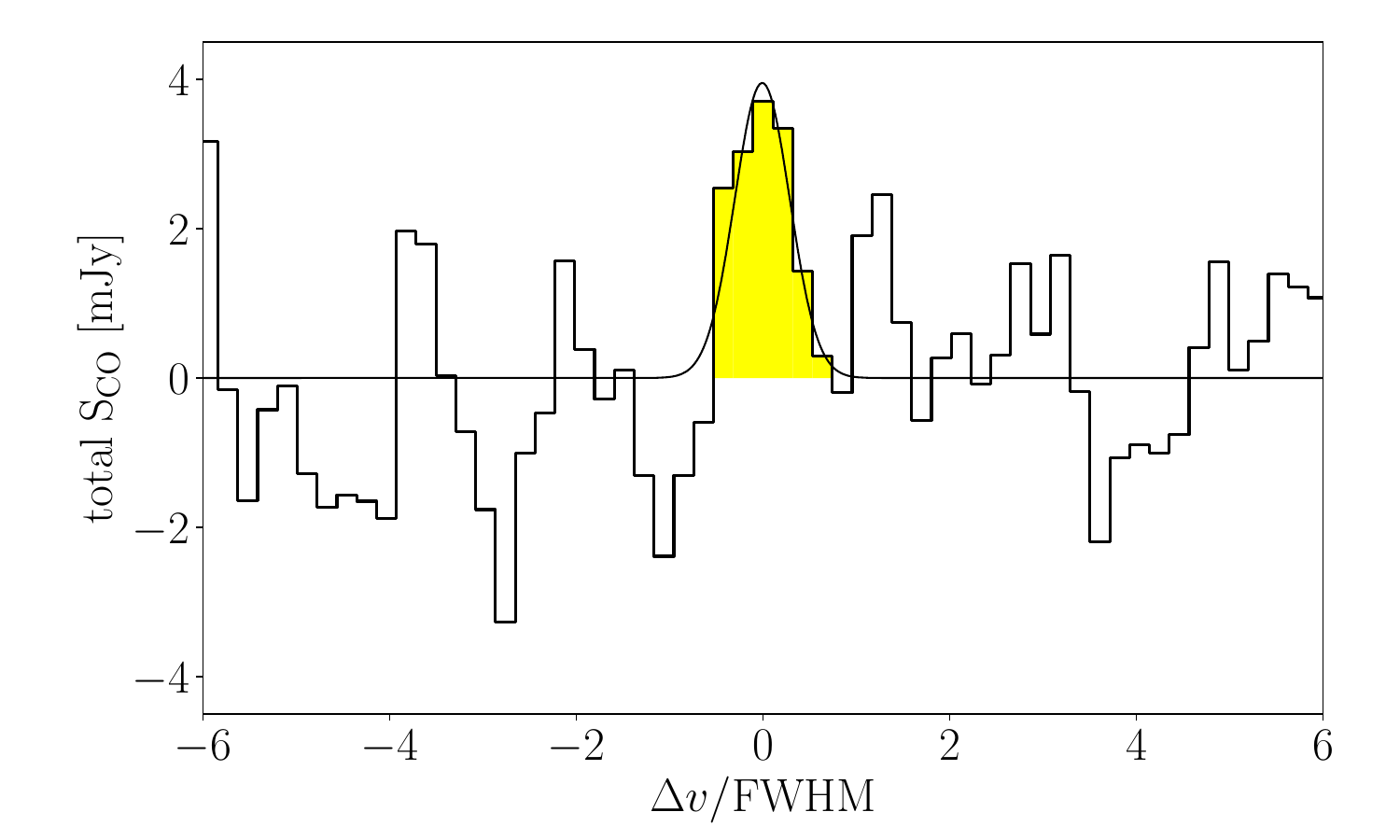}
			}
		\put(14,53){\tiny \scalebox{1.2}{all without 71}}
	\end{picture}
	\\
    \vspace{0.4cm}
	\begin{picture}(100,60)
		\put(-3,0){
		\includegraphics[width=\linewidth,trim={0.cm 0cm 0cm 0cm},clip]{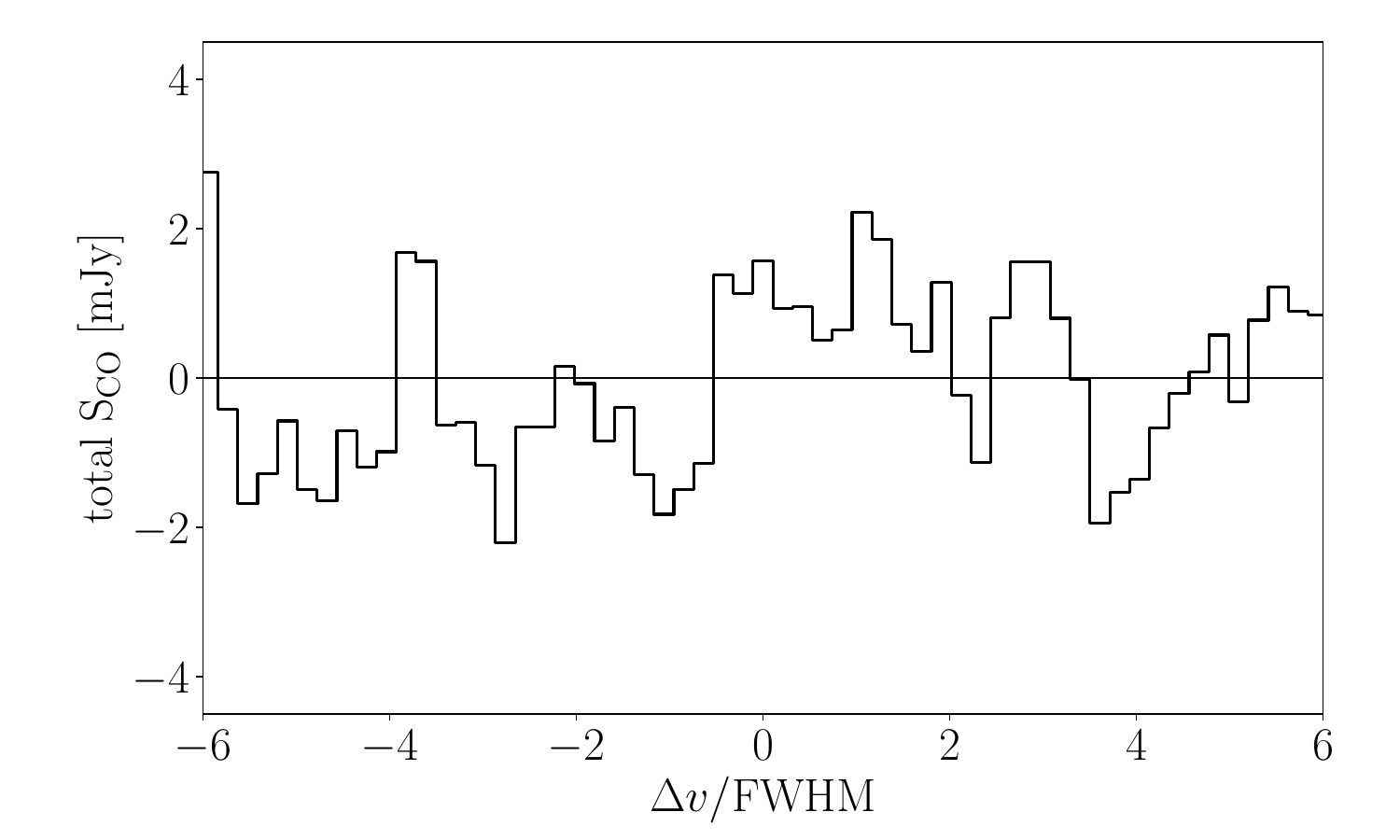}
		}
		\put(14,53){\tiny \scalebox{1.2}{non-detections}}
	\end{picture}
	\caption{Stacked NOEMA CO(2-1) spectra over all galaxies in the primary beam (top panel), over all galaxies except \#71 (middle panel), and over the non-detections (bottom panel). We define the velocities $\Delta \varv$ with respect to the stellar velocities (the \oii\ velocities for \#69 and \#78) and scale them to the \oii\ FWHM (to a fixed $118~\rm km s^{-1}$ for \#57, \#61, \#93, and \#123 in the absence of \oii\ detections). We weighted the fluxes by the NOEMA beam power before stacking. The corresponding signals yield $0.89\ \rm Jy\ km\ s^{-1}$ with a $\rm S/N = 4.7$ for all galaxies, $0.63\ \rm Jy\ km\ s^{-1}$ with a $\rm S/N = 3.9$ for all galaxies except \#71, and a 3$\sigma$ upper limit of $0.40 \ \rm Jy\ km\ s^{-1}$ for the non-detections.}
	\label{fig:stacks}
\end{figure}

For galaxies whose CO(2-1) line is undetected, we used the  MUSE \oii\ FWHM to evaluate upper limits on the integrated line flux.  
Indeed, since the targeted galaxies are typical MS star-forming galaxies, obscuration is not extreme, and both \oii\ and CO should trace similar star-forming regions and velocity fields. Observational studies have shown that \oii\ and CO line widths are comparable \citep[e.g.,][]{Freundlich2013, Puglisi2021}, and this is also the case for the four detected galaxies (cf. Tables~\ref{table:properties} and \ref{table:CO}). 
We evaluated the noise within the FWHM, $\sigma_{\rm FWHM} = \sigma_{\rm RMS} \sqrt{{\rm FWHM}\times \Delta \varv}$, where $\sigma_{\rm RMS}$ is the RMS noise and $\Delta \varv$ is the width of the velocity channels over which it is evaluated, and defined the $3\sigma$ integrated line flux upper limits as $F_{\rm upper}= 3\times \sigma_{\rm FWHM}$,  which correspond to fluxes that should have been detected at $\rm S/N>3$  with a 50\% probability \citep[e.g.,][]{Masci2011}. For the four galaxies undetected in \oii\ (\#57, \#61, \#93, and \#123), we used a fixed FWHM of $118~\rm km s^{-1}$, which corresponds to the median \oii\ dispersion for galaxies with $\log (M_\star/{\rm M_\odot}) <10$.
The beam size is $2.6\times 1.1~\rm arcsec^2$, and the experimental RMS noise is $\sigma_{\rm RMS}= 0.26 ~\rm mJy$ per $\Delta \varv= 44.86~\rm km s^{-1}$ wide channel. 
We also rescaled the upper limit fluxes by the beam power. 

To assess the total and average molecular gas masses in the galaxies, we further stacked the CO spectra at the optical positions of the galaxies in the NOEMA beam, shifting them by the stellar velocity, normalizing the velocities by the MUSE \oii\ FWHM, and weighting by the NOEMA beam power. 
For the four galaxies undetected in \oii, we again use a fixed FWHM of $118~\rm km s^{-1}$ instead of the \oii\ FWHM, while for the two lowest-mass galaxies without \texttt{pPXF} fits we use the \oii\ velocity instead of the stellar velocity.
Fig.~\ref{fig:stacks} shows the resulting stacked spectra for all the galaxies, for all galaxies except \#71, and for the non-detected subsample. We singled out \#71 in the analysis because it is an AGN, which could contaminate the estimated SFR and, consequently, the offset from the MS and the expected molecular gas mass.
For all galaxies in the NOEMA beam, we obtain a signal of 
$0.89 \pm 0.19~ \rm Jy~km~s^{-1}$ with an $\rm S/N = 4.7$; 
for all  galaxies except \#71, we obtain 
$0.63 \pm 0.16~ \rm Jy~km~s^{-1}$ with an $\rm S/N = 3.9$; 
for the non-detections we obtain no stacked detection but derived a 3$\sigma$ upper limit of $0.40 ~\rm Jy~km~s^{-1}$ from the RMS noise ($1.55~\rm mJy$ in channels of 0.21 times the FWHM in the bottom panel of Fig.~\ref{fig:stacks}). 
We stress that this procedure involves no free parameter since both the central velocities and the line widths are constrained.
The relative uncertainties on the stacked CO line fluxes are $\rm 1/(S/N) = 21\%$ for all galaxies and $26\%$ without \#71.
Finally, we searched for dust continuum detection using the full bandwidth of the NOEMA observations, without success.

\subsection{Galaxy molecular gas masses}

From the integrated line fluxes or their upper limits, we derived intrinsic CO(2-1) luminosities, 
\begin{equation}
	\label{eq:luminosity}
	\left(\frac{L^\prime_{{\rm CO(2-1)}}}{\mathrm{K~km~s}^{-1}\mathrm{~pc}^2}\right) = \frac{3.25 \times 10^7}{1+z} 
	\left(\frac{F_{{\rm CO(2-1)}}}{\mathrm{Jy~km~s}^{-1}}\right)
	\left(\frac{\nu_{\rm rest}}{\mathrm{GHz}}\right)^{-2} \left(\frac{D_\mathrm{L}}{\mathrm{Mpc}}\right)^2,
\end{equation}
where $\nu_{\rm rest}$ is the rest-frame frequency and $D_{\rm L}$ is the luminosity distance \citep{Solomon1997}, and the molecular gas masses, 
\begin{equation}
	M_{\rm gas} = \alpha_{\rm CO} L_{{\rm CO}(2-1)}^\prime / r_{\rm 21},
\end{equation}
where $\alpha_{\rm CO}$ is the CO(1-0) luminosity-to-molecular-gas-mass conversion factor and $r_{21}=L_{{\rm CO}(2-1)}^\prime/L_{\rm CO(1-0)}^\prime$ is the corresponding line ratio. 
We adopted the PHIBSS2 methodology \citep{Genzel2015, Tacconi2018, Freundlich2019} to allow comparison with their results and took the conversion factor as the geometric mean of the recipes by \cite{Bolatto2013} and \cite{Genzel2012}, which include a metallicity dependence given by 
\begin{equation}
	\alpha_{\rm CO} = \alpha_G \sqrt{0.67 \times \exp(0.36 \times 10^{8.67-\log{Z}}) \times 10^{-1.27\times (\log{Z}-8.67)}},
\end{equation}
where $\alpha_G = 4.36 \pm 0.9 \rm~ M_\odot/(K~km~s^{-1} pc^{2})$ is the Galactic conversion factor and $\log{Z} = 12+\log({\rm O/H})$ is the metallicity on the \cite{Pettini2004} scale estimated from the mass--metallicity relation, $\log{Z} = 8.74 - 0.087 \times (\log({M_{\star}/{\rm M_\odot}})-b)^2$ with $b = 10.4 + 4.46 \times \log(1+z)-1.78 \times (\log(1+z))^2$ \citep[cf.][and references therein]{Genzel2015}. The resulting values of $\alpha_{\rm CO}$ range between 3.8 and 8.7, except for the very low-mass galaxies \#69 and \#78, for which it reaches 28.4 and 42.3, respectively\footnote{These high values of $\alpha_{\rm CO}$ affect the corresponding upper limits, but not the stacks, for which we consider the median $\alpha_{\rm CO}$.}.
Still following the PHIBSS2 methodology, we assume $r_{21} = 0.77$, as suggested by observations in low- and high-redshift star-forming galaxies  \citep[e.g.,][]{Leroy2009, Dannerbauer2009, Aravena2010, Papadopoulos2012b, Bothwell2013, Daddi2015, Boogaard2020, Prajapati2026} and analytical models of turbulent, clumpy gas disks \citep[e.g.,][]{Vollmer2017, Vollmer2025}. 

\begin{figure}
	\centering
	\includegraphics[width=1\linewidth,trim={0.5cm 0.2cm 1cm 1.5cm},clip]{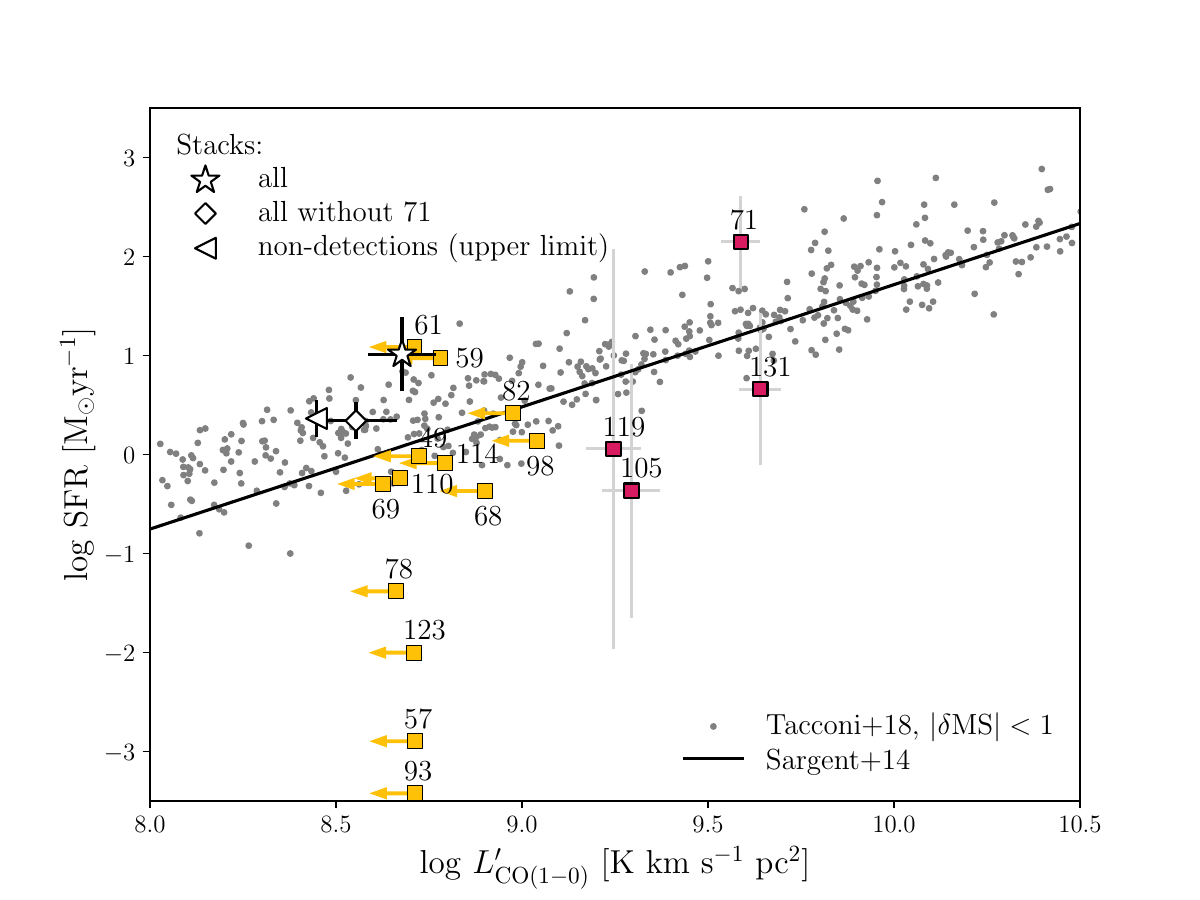}
	\caption{
		Observed relation between the SFR and the intrinsic CO(1-0) luminosity for the sample galaxies (red and orange squares, indicating detections and upper limits, respectively) and for the stacked measurements averaged over all sample galaxies, all galaxies except \#71, and the non-detections (open symbols with error bars). The plot also shows the \citet{Tacconi2018} sample of star-forming galaxies within 1 dex of the MS line up to $z=3.4$ (grey  points). These points are compared to the \citet{Sargent2014} fitting function, whose dispersion is $\sim 0.2 \rm ~dex$. 
        Error bars represent the uncertainties in the SED-based SFRs and the CO line intensities, notably neglecting the uncertainty in the line ratio $r_{21}$. 
        }
	\label{fig:SFR_vs_LCO10}
\end{figure}

Figure ~\ref{fig:SFR_vs_LCO10} shows the correlation between the intrinsic CO(1-0) luminosity, directly derived from the observations assuming $r_{21}=0.77$, and the SED-based SFR. We compared this correlation against two benchmarks: the larger \citet{Tacconi2018} sample, which uses the same assumptions as we did to convert the observed CO(2-1) luminosities into CO(1-0) luminosities and molecular gas masses, and the fitting function proposed by \citet{Sargent2014}. The \citet{Tacconi2018} sample comprises 650 CO molecular gas measurements spanning redshifts from $z=0$ to $3.4$, from which we specifically selected those that fall within 1 dex of the MS line. The \citet{Sargent2014} fitting function is based on a previous compilation including 131 of these measurements. 
Our measurements and upper limits in the CGr30 group seem to be biased toward the upper part of the scatter, as notably shown by the positions of the stacked measurements. This may indicate a dearth of molecular gas in the group environment, as well as an increase in the star formation efficiency. 
In particular, the stack of all 17 galaxies falls left of the \citet{Sargent2014} line by 0.75 dex in $L_{\rm CO(2-1)}$, indicating a molecular gas content that is, on average, only 18\% of the benchmark value. The non-detection stack falls left of the \citet{Sargent2014} line by 0.46 dex, corresponding to 35\% of the benchmark value. 
We note that, although the stack of 17 galaxies from the \citet{Tacconi2018} sample would be unlikely to lie that far from the relation, the galaxies in the present sample are from the same group and could thus all share a similar behavior without necessarily being outliers. 
As galaxy \#71 is an AGN, its SFR may also be contaminated and hence overestimated. A lower SFR may bring the corresponding data point closer to the \citet{Sargent2014} line, so we also considered a stack without this galaxy. In this case, $L_{\rm CO(2-1)}$ only falls short by 0.33 dex, corresponding to 47\% of the benchmark value. 
In Appendix~\ref{appendix:SFR_OII}, we further analyze the \oii\ SFRs rather than the SED-based SFRs, notably because the \oii\ SFR of galaxy \#71 is significantly below the SED-based estimate (cf. Tables~\ref{table:properties} and \ref{table:OII}). Although the individual points differ, the overall results from the stacked measurements are very similar.

\subsection{Molecular gas mass fractions and depletion times}

Molecular gas masses, upper limits, and stacked measurements enable us to derive molecular gas-to-stellar mass ratios, $\mu_{\rm gas}=M_{\rm gas}/M_\star$, and depletion times ,$t_{\rm depl}=M_{\rm gas}/{\rm SFR}$. Table~\ref{table:CO} lists the individual values and highlights that galaxy \#71 has a particularly low depletion time because of its SFR. The stacked measurements yield $\mu_{\rm gas}=0.10\substack{+0.07\\-0.04}$ for both all galaxies and without \#71, and an upper limit of $0.11$ for the non-detections;  $t_{\rm depl}=0.22\substack{+0.50\\-0.17}~\rm Gyr$ for all galaxies, $0.80\substack{+1.72\\-0.73}~\rm Gyr$ without \#71, and an upper limit of $0.65~\rm Gyr$ for the non-detections. These values can be compared to the median values found by \citet{Freundlich2019} between $z=0.5-0.8$, namely $\widetilde{\mu_{\rm gas}}=0.28\pm 0.04$ and $\widetilde{t_{\rm depl}}= 0.84\pm 0.07\rm ~Gyr$. While the values of $\mu_{\rm gas}$ consistently point toward a lower molecular gas fraction than in the comparison sample of typical star-forming galaxies from \citet{Freundlich2019}, more precisely $\sim36\%$ lower, the uncertainties in SFRs make the conclusion less straightforward for the depletion time. Furthermore, both $\mu_{\rm gas}$ and $t_{\rm depl}$ depend on the stellar mass and the offset from the MS, whose influence needs to be normalized to effectively compare the molecular gas reservoirs between CGr30 galaxies and typical star-forming galaxies. 

\begin{figure}
	\centering
	\includegraphics[width=1\linewidth,trim={0.5cm 0.2cm 1cm 1.5cm},clip]{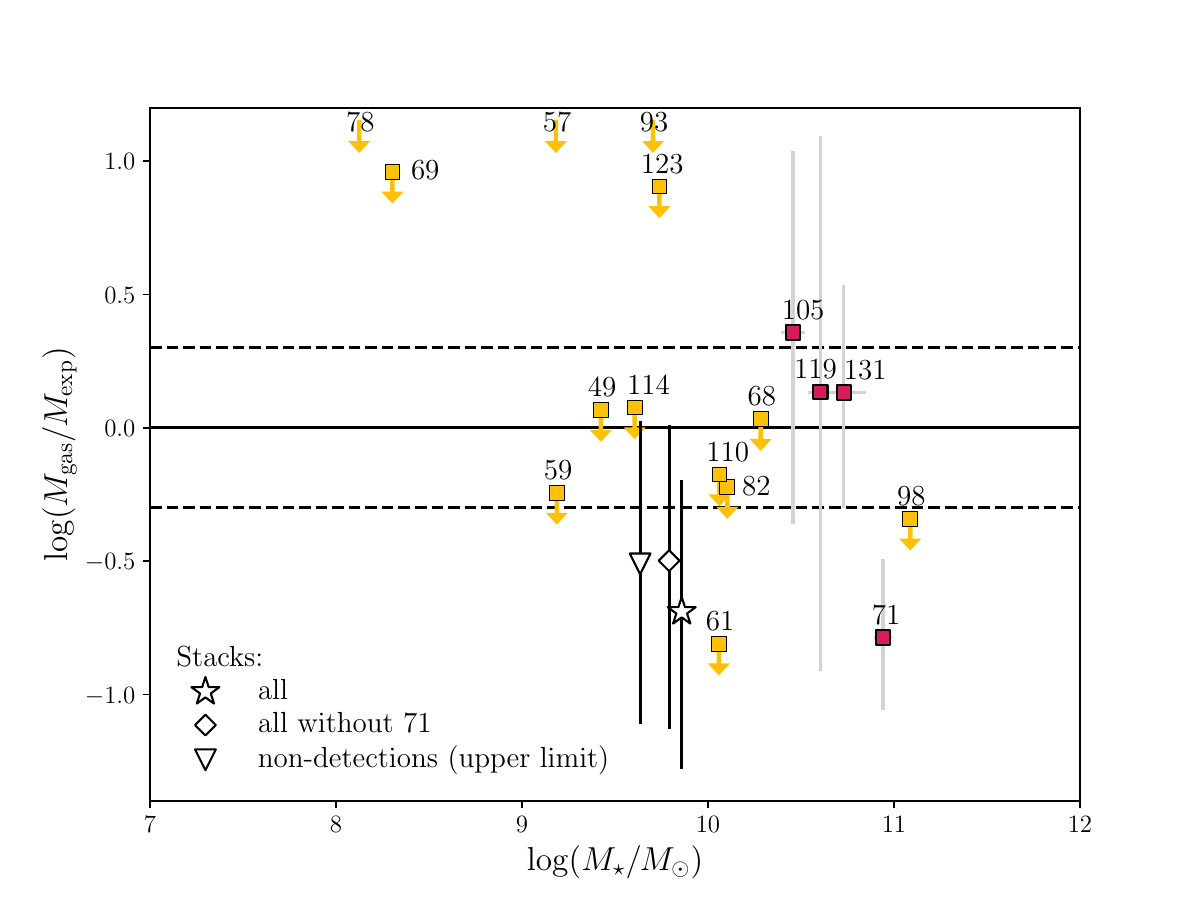}
	\caption{
		Molecular gas mass divided by that expected from the \citet{Tacconi2020} molecular gas mass fraction scaling relation as a function of stellar mass. For comparison with the individual data points and upper limits, the dashed line indicates the scatter of the scaling relation.\textsuperscript{\ref{footnote_scatter}} On average, the stacked molecular gas mass for all galaxies in the sample amounts to $0.20\protect\substack{+0.43\protect\\-0.15}$ of the expected value from the scaling relation. Excluding \#71, the stacked molecular gas mass amounts to  $0.32\protect\substack{+0.71\protect\\-0.24}$. For the non-detections the upper limit  yields $0.31\protect\substack{+0.75\protect\\-0.23}$. 
	}
	\label{fig:mutdepl_vs_DMS}
\end{figure}

To do so, Fig.~\ref{fig:mutdepl_vs_DMS} compares the molecular gas masses ($M_{\rm gas}$) of the sample galaxies to those expected ($M_{\rm exp}$) from the stellar masses and the \citet{Tacconi2020} molecular gas-to-stellar mass ratio scaling relation, which expresses the molecular gas-to-stellar mass ratio expected for typical star-forming galaxies as a function of redshift, stellar mass, and offset from the redshift-dependent MS, 
\begin{equation}
\log \mu_{\rm exp} = A + B\ (\log(1+z)-F)^2 + C\  \delta {\rm MS} + D\ (\log M_\star-10.7),
\label{eq:Tacconi}
\end{equation}
where $A=0.06$, $B=-3.33$, $C=0.51$, $D=-0.41$, $F=0.65$, and $\delta {\rm MS} = {\rm sSFR}/{\rm sSFR}(MS; z, M_\star)$. 
This relation, which builds upon those of \cite{Genzel2015}, \cite{Scoville2017}, and \cite{Tacconi2018}, was obtained for a sample of 2052 star-forming galaxies between $z=0$ and $5.3$, $\log M_\star=9$ and $12.2$, and $\log(\delta {\rm MS})=-1.5$ and $+3.75$. The molecular gas measurements stem from CO for 858 galaxies, far-IR dust measurements for 724 galaxies, and $\sim$1mm dust measurements for 470 galaxies. 
The relation captures the sharp decline of the molecular gas fraction since cosmic noon, but also quantify how the gas fraction is higher above the MS and lower below it. 
Although the samples from which the relations stem are diverse, most of them do not include any environmental selection. 
For individual and stacked measurements, we estimated the median $M_{\rm gas}/M_{\rm exp}$ and their 1$\sigma$ uncertainties through Monte Carlo sampling. For $M_{\rm gas}$, we considered the uncertainty in the CO(2-1) flux measurement and a 0.2 dex uncertainty in the molecular gas conversion factor. For the expected $M_{\rm exp}$, 
we propagated the uncertainties in $\log{\rm SFR}$ and $\log M_{\star}$ first in ${\rm SFR({\rm MS}; z, M_\star)}$, then in $\delta{\rm MS}={\rm SFR}/{\rm SFR}({\rm MS})$, and finally in $\mu_{\rm exp}(z,M_\star,\delta{\rm MS})$ and the expected molecular gas mass $\mu_{\rm exp}  M_\star$. We also accounted for an additional 0.2 dex scatter in ${\rm SFR}({\rm MS})$ and a 0.3 dex scatter in $\mu_{\rm exp}$\footnote{Since the 0.3 dex scatter in $\mu_{\rm exp}$ is accounted for in their error bars, the stacked data points in Fig.~\ref{fig:mutdepl_vs_DMS} should be compared to the solid line $y=0$ and not to the dashed lines that highlight the scatter in the relation for individual galaxies.\label{footnote_scatter}}.

While individual galaxies \#105, 119, and 131 fall within the expected scatter, the stacked measurements indicate an average trend for CGr30 galaxies to have lower molecular gas masses than the typical star-forming galaxies underlying the \citet{Tacconi2020} scaling relations. 
The stacked molecular gas mass of all galaxies is $0.20\substack{+0.43\\-0.15}$ times the expected value for typical star-forming galaxies; without \#71, it is $0.32\substack{+71\\-0.24}$; and for the non-detections, the upper limit is $0.31\substack{+0.75\\-0.23}$. These three values correspond to distances from the expected values of $1.40$, $0.98$, and $0.95\sigma$, respectively. The corresponding probabilities that the molecular gas masses are below the expected values are $92\%$, $84\%$, and $83\%$, respectively. We note, however, that the error bars remain relatively large, in part due to the uncertainties in the SED-based SFR estimates.
Nevertheless, Appendix~\ref{appendix:SFR_OII} shows that we obtain similar results with the \oii\ SFRs instead of the SED-based SFRs, which strengthens the evidence for lower molecular gas masses than in typical star-forming galaxies. 
Since we evaluated $M_{\rm exp}$ from the scaling relation for the molecular-to-stellar mass ratio, $M_{\rm gas}/M_{\rm exp}$ is also a measure of $\mu_{\rm gas}/\mu_{\rm exp}$. As shown in Appendix \ref{appendix:tdepl}, if we instead evaluate the expected molecular gas masses from the scaling relation for the depletion time and the SFRs (which are more uncertain than the stellar masses), we do not find the same discrepancy for the stacked measurements as in Fig.~\ref{fig:mutdepl_vs_DMS}, indicating that the average depletion time for CGr30 galaxies is in line with those of the typical star-forming galaxies underlying the \citet{Tacconi2020} scaling relations.
Our measurements thus seem to indicate that environmental processes deplete molecular gas reservoirs in CGr30 galaxies, typically down to $20-40\%$ of the values observed in normal star-forming galaxies. Possible processes responsible for this dearth include tidal interactions, ram-pressure stripping, and strangulation of the gas supply to the galaxies. 
In CGr30, the gas expelled from the galaxies may be partly responsible for the diffuse \oii-emitting, non-pristine gas detected by MUSE. 
In Appendix \ref{appendix:cw}, we present a further attempt to correlate the molecular gas fraction with the distance to cosmic web filaments, but significantly larger samples in well-studied fields such as COSMOS would be necessary to quantify any such correlation.

\subsection{Total and intragroup molecular gas mass}

\begin{figure}
	\centering
	\begin{picture}(100,60)
		\put(-4,0){
			\includegraphics[width=\linewidth,trim={0.cm 0cm 0cm 0cm},clip]{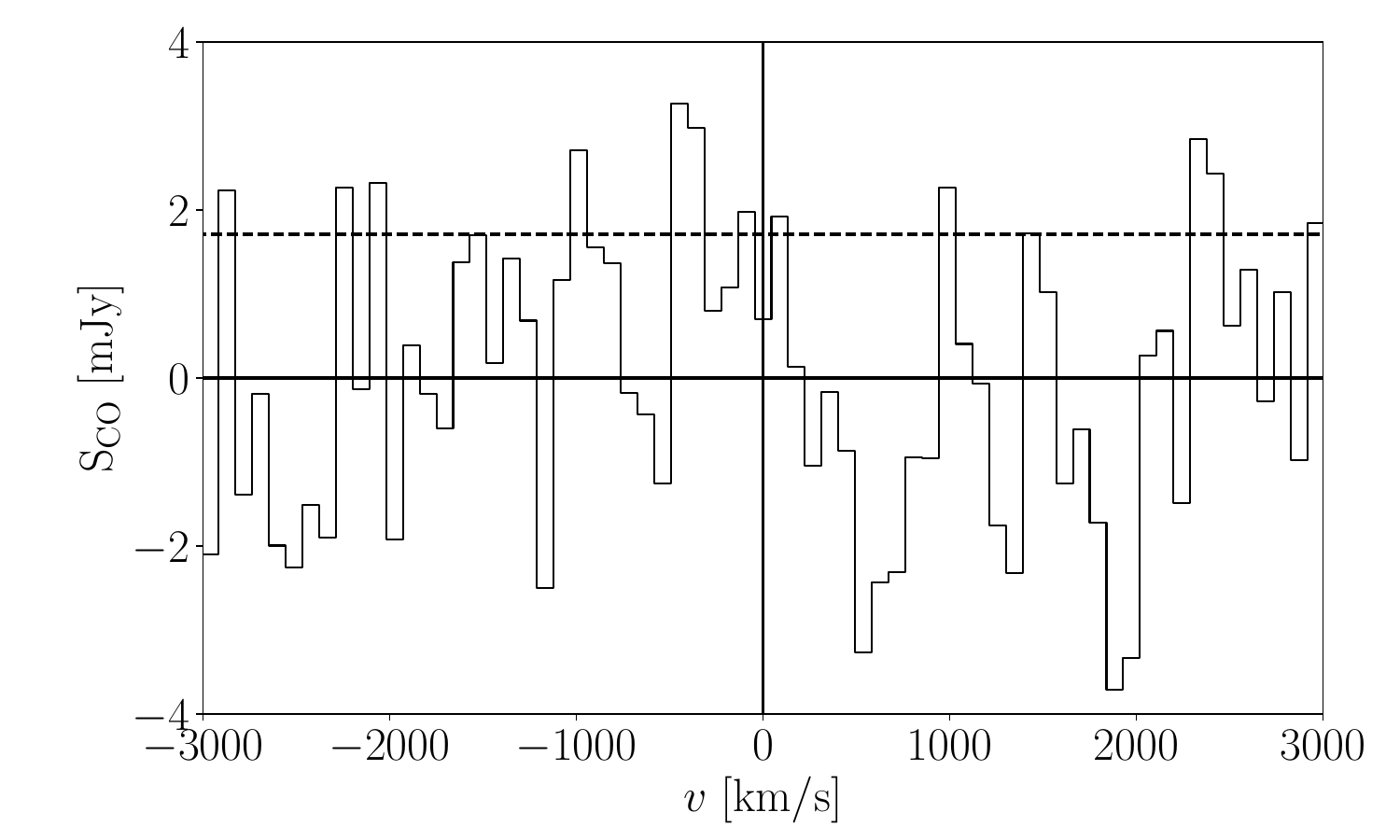}
		}
		\put(13,53){\tiny \scalebox{1.2}{30m}}
	\end{picture}
	\caption{
		Spectrum of the CO(2-1) emission obtained with WILMA using the 30m telescope, yielding no detection. The horizontal dashed line indicates the RMS noise of $1.71~\rm mJy$ over  $\rm 89.74 ~km s^{-1}$ velocity channels. 
	}
	\label{fig:30m}
\end{figure}
\begin{figure}
	\centering
	\begin{picture}(100,60)
	\put(-4,0){
		\includegraphics[width=\linewidth,trim={0cm 0cm 0cm 0cm},clip]{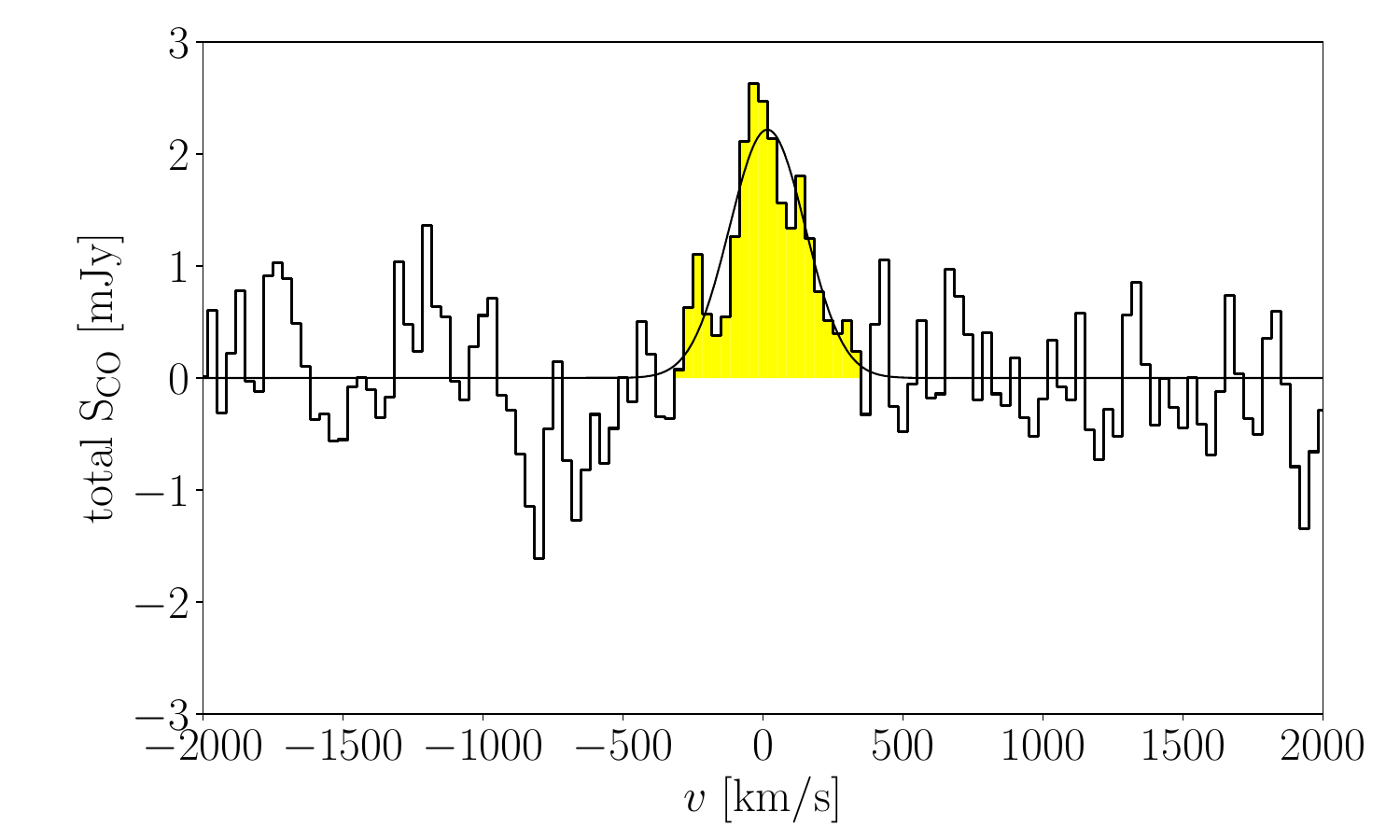}
	}
	\put(13,53){\tiny \scalebox{1.2}{30m beam }}
\end{picture}
	\caption{
	Stacked NOEMA CO(2-1) spectrum weighted by the 30m beam power. Velocities are not shifted by the stellar velocity, unlike in Fig~\ref{fig:stacks}, and 
    we weighted the fluxes by the NOEMA beam power before stacking. The resulting signal yields $0.75\ \rm Jy\ km\ s^{-1}$ with a $\rm S/N = 8.3$. 
	}
\label{fig:noema_30m_stack}
\end{figure}

The complementary 30m CO(2-1) observations were aimed at assessing the overall molecular gas content in the CGr30 group, including possible extended structures in the intragroup medium. As shown in Fig~\ref{fig:30m}, the observations show no signal, but it is still possible to establish an upper limit. 
Given the median \oii\ FWHM of the individual galaxies and the group-scale velocity dispersion derived from the stellar velocities, the expected FWHM of the overall signal is $\sim 600~\rm km~s^{-1}$. This value also matches the velocity span previously observed for the diffuse \oii\ component in this system \citep{Epinat2018}. Given the RMS noise ($1.7 ~\rm mJy$ over $\rm 89.74 ~km s^{-1}$ velocity channels), we evaluated the noise within this FWHM at $\sigma_{\rm FWHM, 30m}=0.40 ~\rm Jy~km~s^{-1}$, which yielded a 3$\sigma$ integrated line flux upper limit of $F_{\rm upper, 30m} = 1.19 \rm ~Jy~km~s^{-1}$. Using the median redshift, luminosity distance, and CO conversion factor, we derive a 30m total molecular gas mass upper limit of $M_{\rm gas, tot}< 5.4 \times 10^{10}~\rm M_\odot$ (or $\log M_{\rm gas, tot}<10.73$).

For comparison with the measurements within galaxies, we stacked the NOEMA spectra, weighted by the 30m beam power and assuming a Gaussian beam with an 18\arcsec\ half-power beam width. Figure~\ref{fig:noema_30m_stack} shows the resulting spectrum, which yields an integrated line flux of $0.75 ~\rm Jy ~km~s^{-1}$ with an $\rm S/N=8.3$. 
Hence, the CO(2-1) line emission exceeding that of the galaxies in the 30m beam should have an integrated intensity lower than $0.37 ~\rm Jy ~km~s^{-1}$. 
Using the median redshift, luminosity distance, and CO conversion factor, this flux translates into an overall molecular gas mass in the intragroup medium below $2\times 10^{10}~\rm M_\odot$, i.e., less than a third of the total gas mass of the large and massive diffuse structure revealed by MUSE, which was estimated to be $\sim 5 \times 10^{10}~\rm M_\odot$ \citep{Epinat2018}.

\section{Conclusions}
\label{section:conclusion}

In this article, we present CO(2-1) molecular gas measurements in an overdense region of a galaxy group at $z\sim0.7$ using both the IRAM NOEMA interferometer and 30m telescope. We selected the target group, COSMOS-Gr30, from a sample of intermediate redshift groups that were observed with MUSE to study environmental quenching. It is the densest group of this sample and it is located at the intersection of large-scale cosmic web filaments. Deep MUSE observations reveal a large, massive structure of ionized gas surrounding eight galaxies of the group, including one active galactic nucleus (AGN) host. This gas is not pristine and was likely expelled from the galaxies through both internal processes, such as star formation and AGN outflows, and external processes in the dense environment, such as ram pressure stripping and tidal interactions. 

Using NOEMA, we detect molecular gas with $\rm S/N>5$ in four individual galaxies and derive upper limits for the other 13 galaxies that lie within the NOEMA primary beam. We also obtain stacked detections for the total molecular gas content in the 17 galaxies. Although some of the group galaxies have typical molecular gas reservoirs, the stacked detections point toward lower molecular gas content in the group galaxies on average, possibly due to environmental processes. Indeed, the stacked detection of all galaxies within the NOEMA beam indicates molecular gas masses that are generally $20-40\%$ of those found in typical star-forming galaxies. This depletion suggests that environmental processes in the group efficiently remove or suppress the molecular gas required for star formation.

Using the 30m telescope, we obtain an upper limit for the total molecular gas emission in the beam, which includes most of the extended ionized gas structure uncovered by MUSE. Together with a stacked detection of the molecular gas within the galaxies in the 30m telescope beam, this allowed us to estimate that the extended gas structure contains less than $2\times 10^{10} \rm ~M_\odot$ molecular gas. This suggests that less than a third of the gas displaced from the galaxies into the diffuse structure is in a cold phase.

This work extends the characterization of molecular gas reservoirs in dense environments toward group scale and beyond the dominant central systems, and highlights the importance of simultaneously probing gas reservoirs within galaxies and the surrounding intragroup medium to capture environmental quenching in action. In the case of CGr30, the depletion of molecular gas within galaxies, combined with the presence of an extended gas structure, provides a consistent picture of environmental quenching. However, these processes require further quantification through deeper observations with NOEMA, the IRAM 30m telescope, and ALMA. Observations with ALMA may also reveal the molecular counterpart of the extended gas structure. Furthermore, more precise SFR measurements are necessary to decrease the uncertainties, and larger samples of target galaxies in groups will allow us to assess group-to-group variance. 
\\

\noindent
   \begin{spacing}{0.81}

    {\tiny \noindent
    \textit{Data availability.}  The spectral data shown in Fig.~\ref{fig:detections} are available in electronic form at the CDS via anonymous ftp to cdsarc.u-strasbg.fr (130.79.128.5) or via \url{http://cdsweb.u-strasbg.fr/cgi-bin/qcat?J/A+A/}. 
    }
    
   \end{spacing}

\begin{acknowledgements}
    We thank the anonymous referee for their careful reading and suggestions.
	This work is based on observations carried out under project S19BV with the IRAM NOEMA Interferometer and projects 110-19 and 173-20 with the IRAM 30m telescope. IRAM is supported by INSU/CNRS (France), MPG (Germany) and IGN (Spain). We are grateful for the excellent support provided by the IRAM staff. We thank Clotilde Laigle for providing the initial catalogs of realizations of the cosmic web in the COSMOS field, and granting us permission to use them as presented in Appendix~\ref{appendix:cw}.
    D.K. and C.M.-L. acknowledge the support by the Deutsche Forschungsgemeinschaft, DFG project number 4548/4-1. 
	R.H.-C. thanks the Max Planck Society for support under the Partner Group project ``The Baryon Cycle in Galaxies'' between the Max Planck for Extraterrestrial Physics and the Universidad de Concepción. L.A.B. acknowledges support from the Dutch Research Council (NWO) under grant VI.Veni.242.055 (\url{https://doi.org/10.61686/LAJVP77714}). R.H-C. also gratefully acknowledges financial support from ANID - MILENIO - NCN2024\_112 and ANID BASAL FB210003. 
\end{acknowledgements}

%

\bibliographystyle{aa} 
\bibliography{Freundlich_CGR30} 

\begin{appendix}





\section{CGr30 within the cosmic web}
\label{appendix:cw}

\begin{figure*}[h]
	\centering
	\includegraphics[width=0.9\linewidth,trim={0cm 0cm 0cm 1.cm},clip]{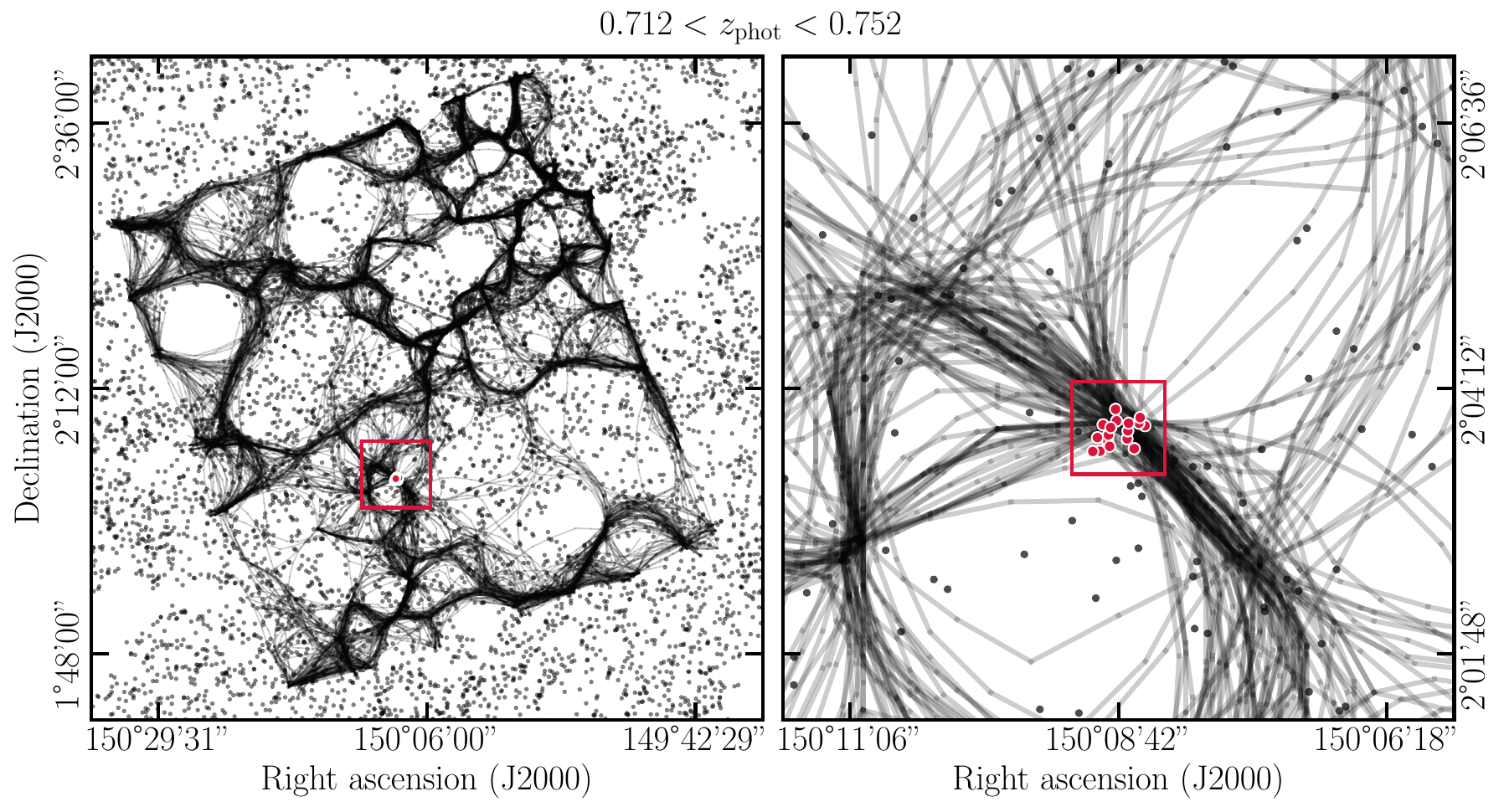}
    \caption{\textit{(Left panel:)} Galaxies in the COSMOS field (grey), with CGr30 members highlighted in red, together with 100 overplotted realisations of the cosmic web (filaments) within the COSMOS-Web footprint, in a redshift slice between $0.712< z_{\rm phot}< 0.752$. 
    The red box indicates the region shown in the right panel. \textit{(Right panel:)} Zoom-in on the main filaments surrounding CGr30. The red box marks the region shown in Fig.~\ref{fig:targetsfield}.}
	\label{fig:cw}
\end{figure*}

Figure~\ref{fig:cw} shows the position of CGr30 in the large-scale environment using two dimensional cosmic web reconstructions from the galaxy position field projected along the line of sight using the publicly available software {\sc DisPerSE} \citep{sousbie2011, sousbie+2011}.
The cosmic web reconstruction follows the methodology of \citet{laigle2025} and \citet{jego2026b}, using all galaxies in the COSMOS field and tomographic slices with a constant comoving thickness set by the photometric redshift uncertainty of low-mass galaxies in the sample.
In COSMOS-Web, the photometric redshift accuracy allows slices with projection depth of $80\,h^{-1}\,\rm Mpc$. Filaments are identified with {\sc DisPerSE} at a $2.5\sigma$ persistence threshold.
We refer the reader to \citet{laigle2025} for details on the cosmic web reconstructions from {\sc DisPerSE} \citep[see also][]{laigle2018}.
We generate 100 Monte Carlo realizations by sampling photometric redshift probability distributions and propagate these into the filament reconstruction and distance measurements.
Mask filling is applied in masked regions by sampling random galaxy positions with an average galaxy number density matching that of the tomographic slice, although the field shown here lies sufficiently far from masked areas so that this correction has a negligible impact.

\begin{figure}[h]
	\centering
	\includegraphics[width=0.9\linewidth,trim={0cm 0cm 0cm 1.cm},clip]{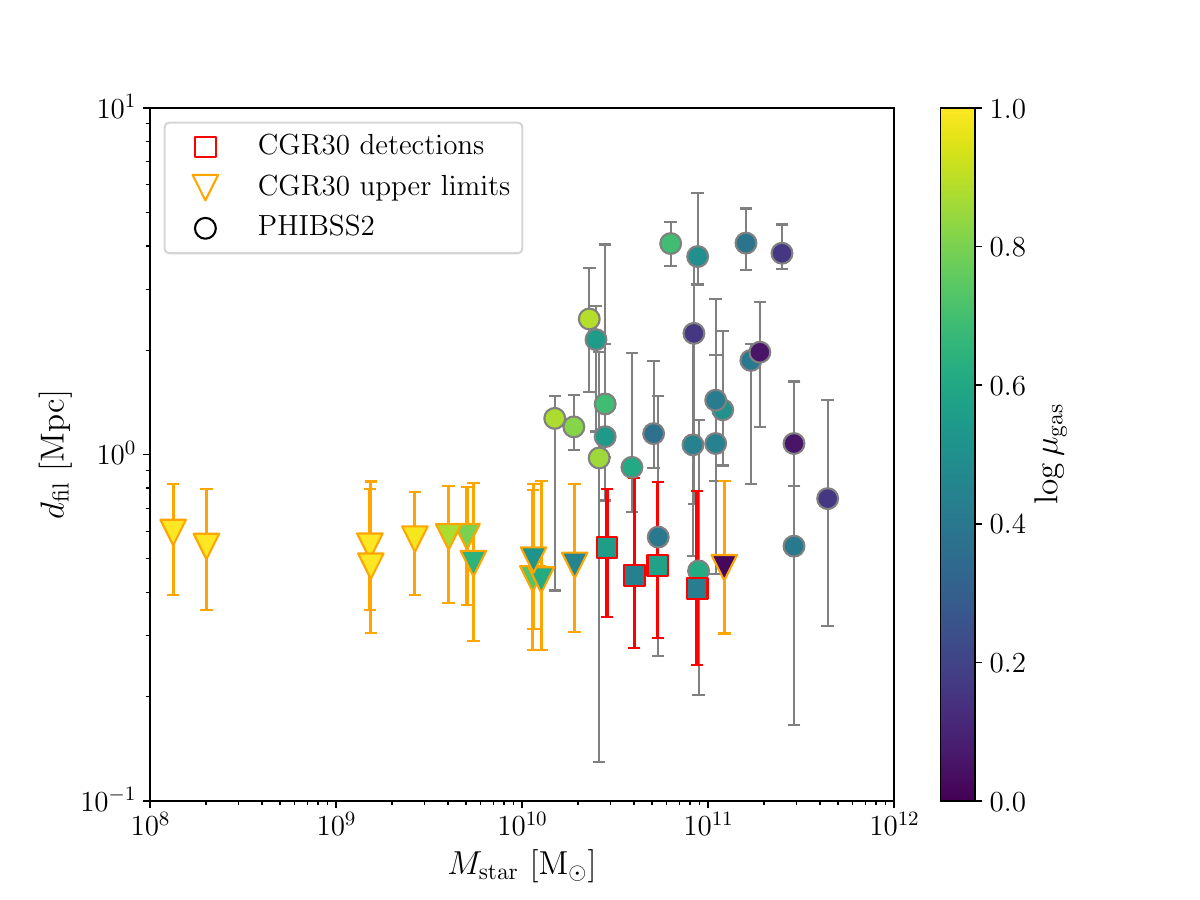}
    \caption{
        Distance to the nearest cosmic web filament, computed as the average over the 100 realisations with Bootstrap errors,  as a function of stellar mass for CGr30 galaxies and PHIBSS2 \cite{Freundlich2019} typical star-forming galaxies between $z=0.5-0.8$ in the COSMOS field, colored according to the molecular gas fraction $\mu_{\rm gas}=M_{\rm gas}/M_\star$. 
    }
	\label{fig:cw_mugas}
\end{figure}

Figure~\ref{fig:cw_mugas} displays the distance to the nearest cosmic filament as a function of stellar mass. We compare CGr30 galaxies with the PHIBSS2 sample  of typical star-forming galaxies at $z=0.5-0.8$ in the COSMOS field from \citet{Freundlich2019}. Although the PHIBSS2 galaxies were selected to span the MS scatter without morphological constraints, their relatively regular appearances suggest they are largely isolated systems. The median distance to the nearest filament is 
$0.48$ Mpc for the CGr30 galaxies and $1.45$ Mpc for the PHIBSS2 control sample. 
As a proof of concept, we explored a possible trend of the molecular gas fraction $\mu_{\rm gas}=M_{\rm gas}/M_\star$ with distance to the nearest filament. Focusing on the mass range between $\log M_\star=10.4-11$ where the four CGr30 detected galaxies lie, we find $\mu_{\rm gas} = 0.23 \pm 0.10$ for the four detected CGr30 galaxies and $0.28 \pm 0.06$ for the PHIBSS2 galaxies, i.e., no significant trend. To assess the link between molecular gas reservoirs in galaxies and their position with respect to the cosmic web, more observations in well-studied fields such as COSMOS would be needed, spanning different environments. This preliminary exploration of the correlation between gas fraction and distance to cosmic filaments thus further underscores the need for significantly larger samples to yield statistically significant results.

\section{Star formation rates from \oii}
\label{appendix:SFR_OII}

\begin{table}
	\caption{\oii\ fluxes and SFRs.
	}             
	\label{table:OII}      
	\centering          
	\footnotesize
\begin{tabular*}{0.48\textwidth}{@{\extracolsep{\fill}} rrr}
    \hline\hline \\[-0.3cm]
    \multicolumn{1}{l}{${\rm ID}$} & 
    \multicolumn{1}{l}{$F({\rm \oii})$} & 
    \multicolumn{1}{l}{$\log {\rm SFR}_{\rm \oii}$}\\
    
    & 
    \multicolumn{1}{l}{$[\rm 10^{-18} erg\ s^{-1}\ cm^{-2}]$} & 
    \multicolumn{1}{l}{$\rm [M_\odot \rm yr^{-1}]$} \\
    \hline \\[-0.3cm]    
49 & $13.36$ & $0.09 \pm 0.30$\\
57 &  & \\
59 & $1.28$ & $0.87 \pm 0.30$\\
61 &  & \\
68 & $24.09$ & $0.17 \pm 0.30$\\
69 & $16.05$ & $-0.47 \pm 0.30$\\
71 & $7.90$ & $1.36 \pm 0.30$\\
78 & $80.77$ & $-0.74 \pm 0.31$\\
82 & $8.84$ & $0.28 \pm 0.30$\\
93 &  & \\
98 & $4.22$ & $1.37 \pm 0.67$\\
105 & $25.20$ & $0.66 \pm 0.30$\\
110 & $3.10$ & $0.01 \pm 0.30$\\
114 & $59.83$ & $-0.45 \pm 0.31$\\
119 & $12.79$ & $0.75 \pm 0.30$\\
123 &  & \\
131 & $14.26$ & $1.43 \pm 0.30$\\
    \hline                  
\end{tabular*}
\tablefoot{The table lists the \oii\ fluxes and the resulting $\rm SFR_{\rm \oii}$ with their uncertainties for the galaxies in the NOEMA beam. Four galaxies (\#57, \#61, \#93, \#123) are not detected in \oii\ with MUSE. }
\end{table}

\begin{figure}
	\centering
    \begin{picture}(100,60)
    \put(-4,0){
		\includegraphics[width=\linewidth,trim={0cm 0cm 0cm 0cm},clip]{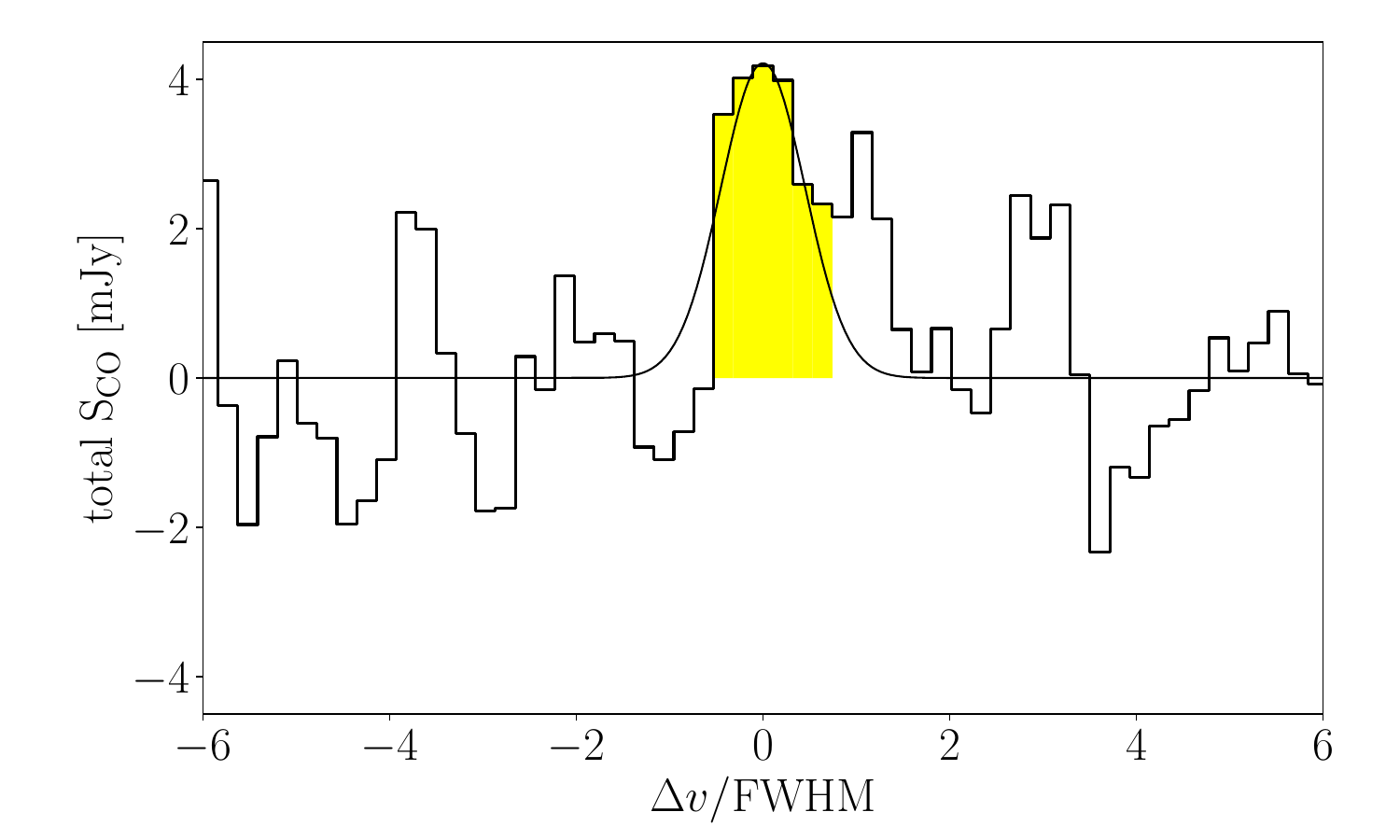}
	}
	\put(13,53){\tiny \scalebox{1.2}{\oii\ SFRs}}
	\end{picture}
	\caption{Stacked NOEMA CO(2-1) spectrum for galaxies with \oii\ SFR measurements, namely without \#57, \#61, \#93, \#123. 
    The corresponding signal yields $1.25\ \rm Jy\ km\ s^{-1}$ with a $\rm S/N = 4.8$. The Gaussian fit was carried out imposing its center to be at zero. 
    }
	\label{fig:stacks_OII}
\end{figure}

To assess the sensitivity of our results to the assumed SFR values and their uncertainties, we also derived SFRs from \oii\ following \citet[][cf. their section 3.3]{Mercier2022}. Namely, we extracted MUSE spectra within apertures of 3\arcsec\ in diameter, subtracted the continuum with \texttt{pPXF}, and fitted the \oii\ doublet with the 1D function of \texttt{Camel}. We corrected the resulting \oii\ luminosities for extinction using the mass-dependent prescription from \cite{Gilbank2010, Gilbank2011} for the intrinsic extinction at the rest-frame H$\alpha$ wavelength and assuming a \cite{Cardelli1989} extinction law with $R_V=3.1$ for the Galactic extinction at the \oii\ wavelength. Finally, we used the expression from \cite{Kennicutt1998b} to convert the  \oii\ luminosities into SFRs. The resulting values are indicated in Table~\ref{table:OII}.

\begin{figure*}
	\centering
	\includegraphics[width=0.495\linewidth,trim={0.5cm 0.2cm 1cm 1.5cm},clip]{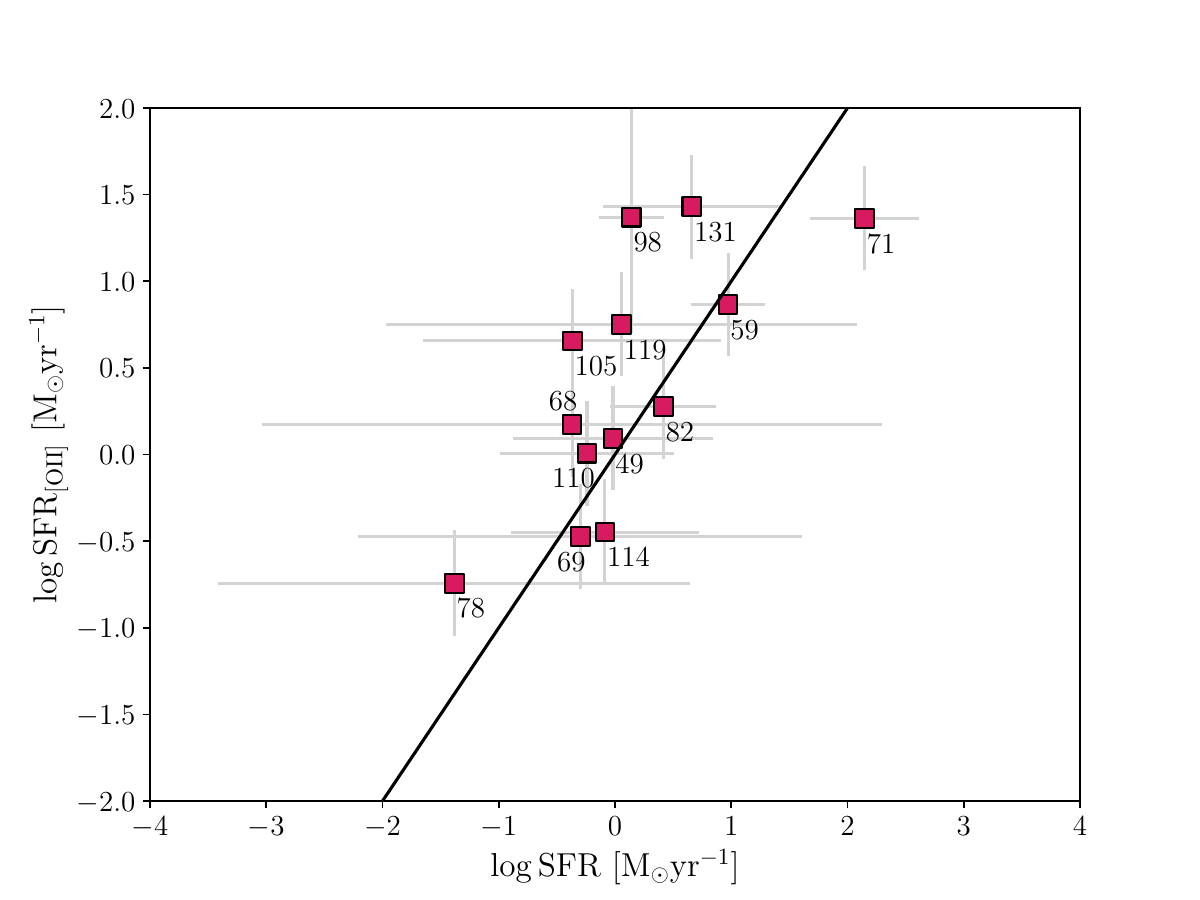}
    \hfill
    \includegraphics[width=0.495\linewidth,trim={0.5cm 0.2cm 1cm 1.5cm},clip]{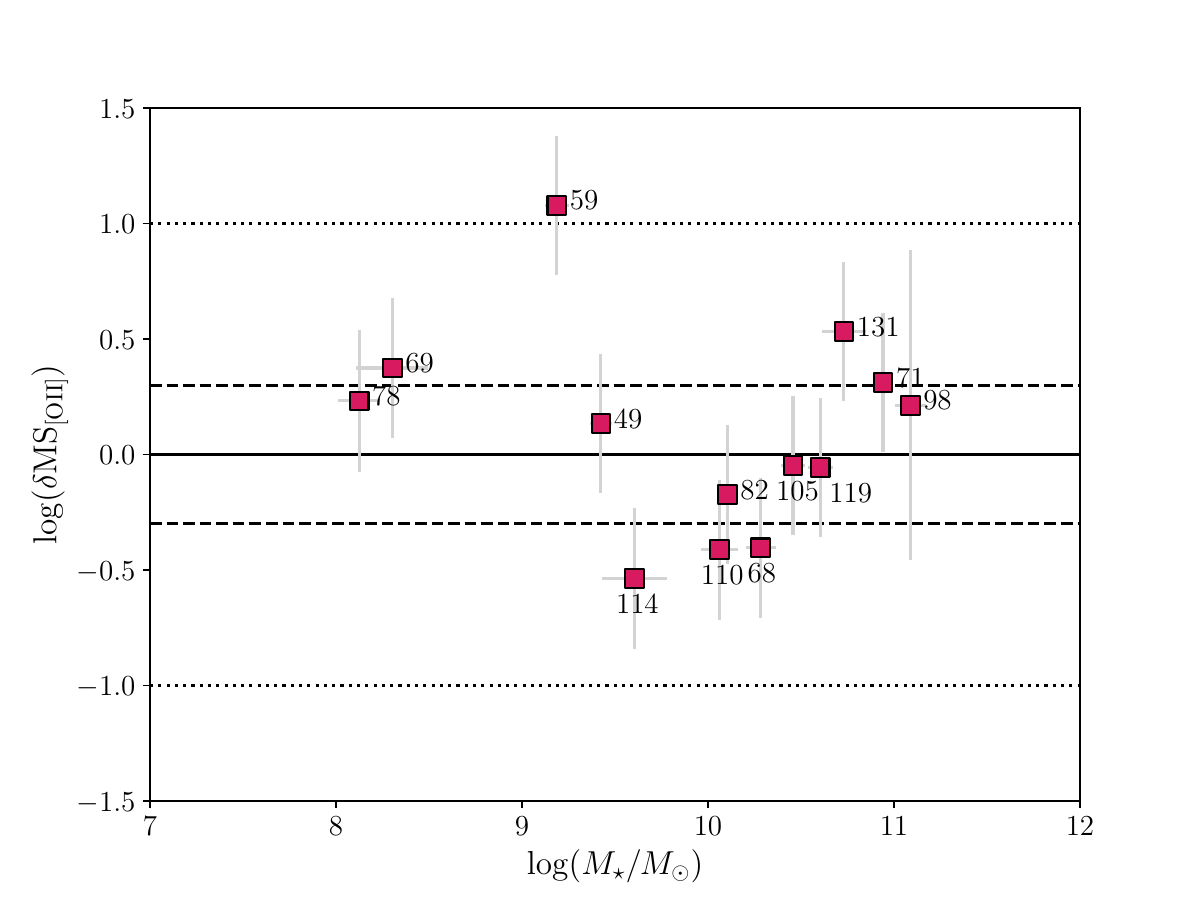}
	\caption{
		\textit{(Left:)} Comparison between the SED-based $\rm SFR$ and $\rm SFR_{\rm \oii}$ from the \oii\ line observed by MUSE. The Pearson correlation coefficient is equal to 0.72. 
        \textit{(Right:)} Position of the galaxies with respect to the MS, as in Fig.~\ref{fig:MS} but using the \oii\ SFRs instead of the SED-based ones and the MS parametrization of \citet{Mercier2022} without downward adjustment. 
	}
	\label{fig:SFR_OII}
\end{figure*}

\begin{figure*}
	\centering
	\includegraphics[width=0.495\linewidth,trim={0.5cm 0.2cm 1cm 1.5cm},clip]{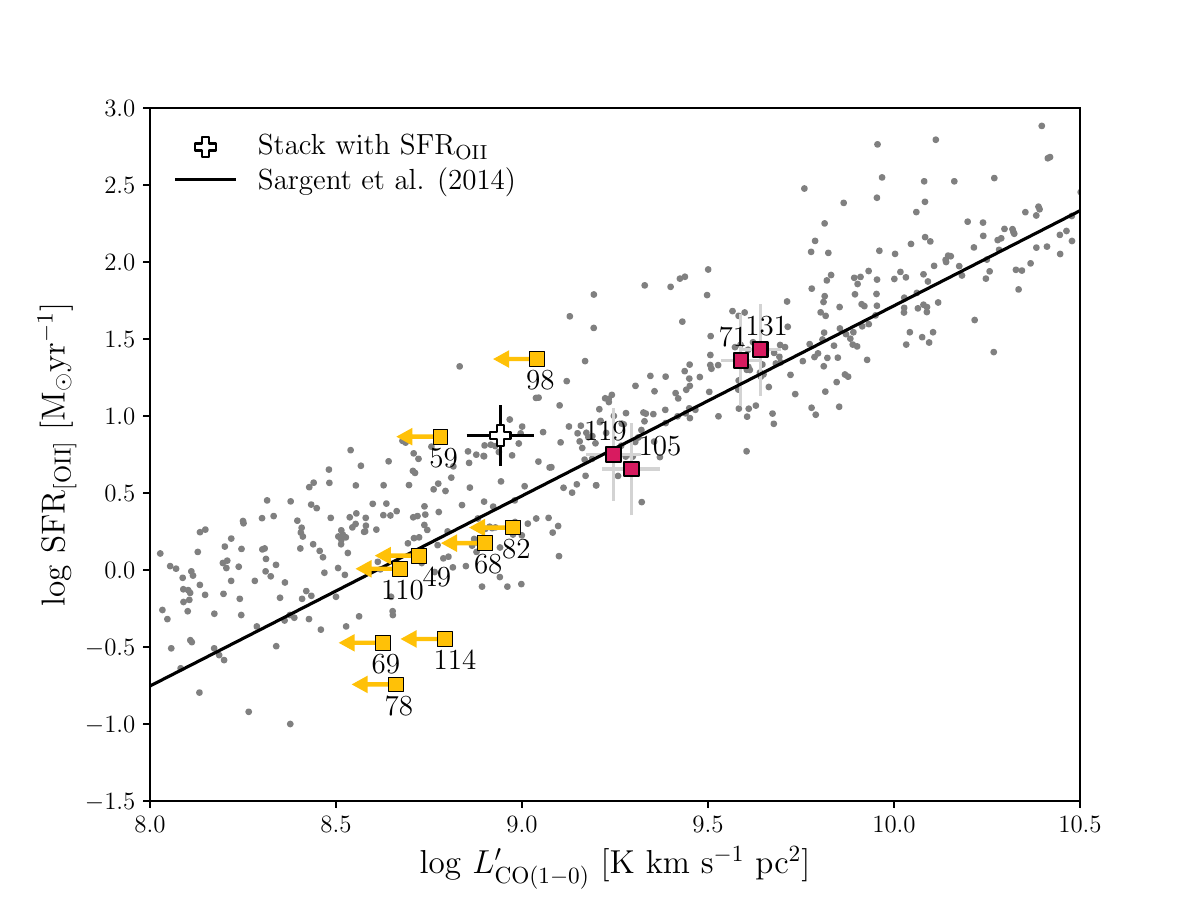}
    \hfill
    \includegraphics[width=0.495\linewidth,trim={0.5cm 0.2cm 1cm 1.5cm},clip]{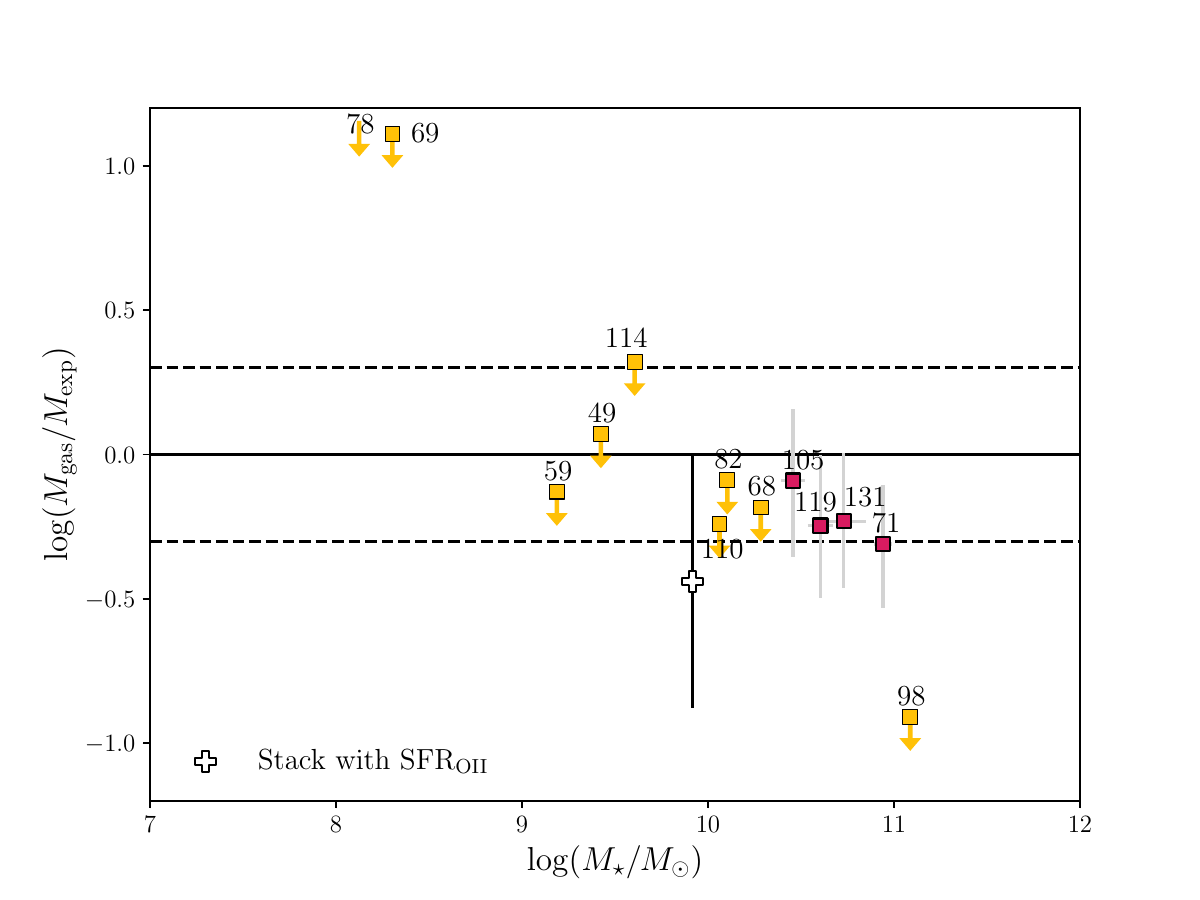}
	\caption{
		\textit{(Left:)} Observed relation between $\rm SFR_{\rm \oii}$ and the intrinsic CO(1-0) luminosity. \textit{(Right:)} Molecular gas mass divided by that expected from the \citet{Tacconi2020} scaling relations as a function of stellar mass, using the  \oii\ SFR. On average, the stacked molecular gas mass corresponds to $0.36\protect\substack{+0.64\protect\\-0.23}$ times the expected value from the scaling relation, and the probability to be below the scaling relation value is 84\%. 
        Detected individual galaxies are highlighted in red.
        }
	\label{fig:Mgas_SFROII}
\end{figure*}

The left panel of Fig.~\ref{fig:SFR_OII} compares the \oii-based $\rm SFR_{\rm \oii}$ with the SED-based values. There is an overall linear trend (the Pearson correlation coefficient is equal to 0.72), but differences for individual galaxies and in particular for \#71, whose $\rm SFR_{\rm \oii}$ is significantly lower than the SED-based SFR. Uncertainties in $\rm SFR_{\rm \oii}$ are assumed to be the combination of the flux uncertainty and a 0.3 dex uncertainty in the conversion factor (which dominates the resulting uncertainty in most cases). 
The right panel shows the resulting position of the galaxies with respect to the MS, assuming the \citet{Mercier2022} parametrization of the MS without downward adjustment (since this downward adjustement precisely corresponds to the average offset between \oii\ SFRs and SED-based SFRs in the MAGIC sample). 
Fig.~\ref{fig:Mgas_SFROII} then compares the CO luminosity and the molecular gas mass associated to individual galaxies and stacks with those expected from scaling relations, as in Figs.~\ref{fig:SFR_vs_LCO10} and \ref{fig:mutdepl_vs_DMS}. Although the individual data points differ from those in these latter figures, the conclusions are similar. In particular, the stacked measurement yields $0.36\substack{+0.64\\-0.23}$ of the expected value for the molecular gas mass.

\section{Star formation histories}
\label{appendix:sfh}

\begin{figure*}
	\centering
    \begin{picture}(100,40)
		\put(0,0){ 
			\includegraphics[width=0.495\linewidth,trim={2.2cm 0.cm 2.cm 1.cm},clip]{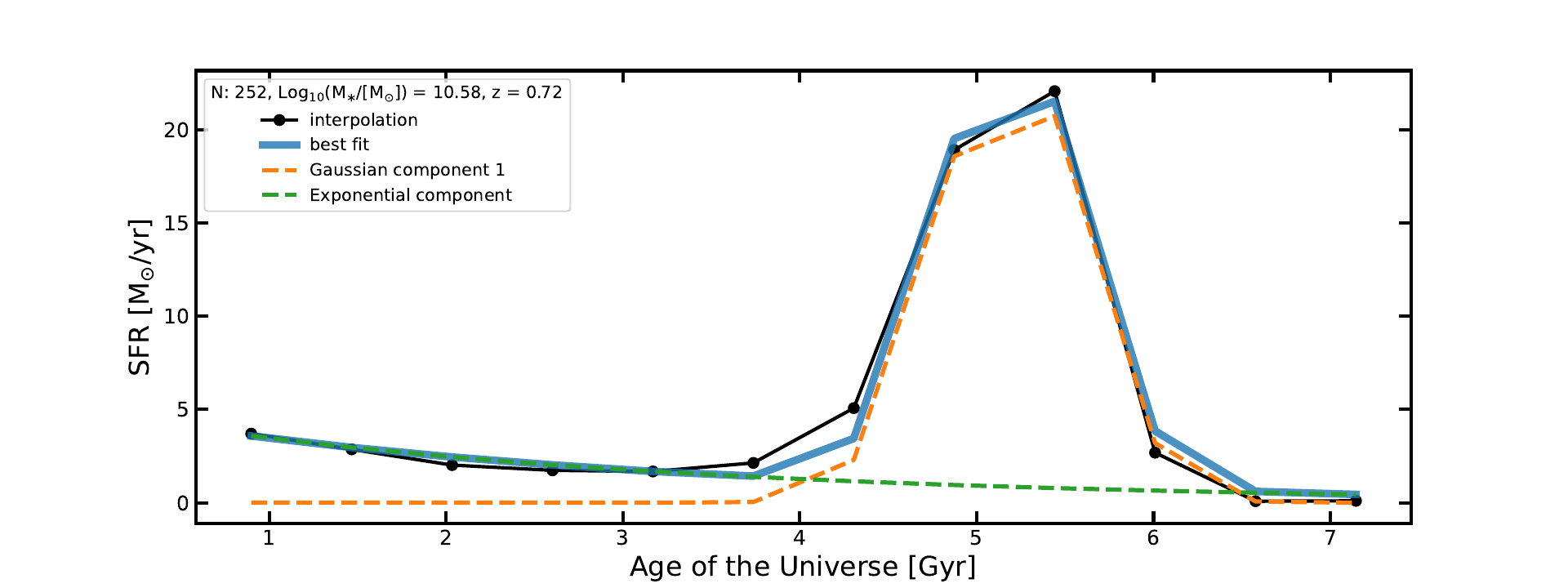}
			}
		\put(45,34){\tiny \scalebox{1.2}{\#71}}
	\end{picture}
    \hfill
    \begin{picture}(100,40)
		\put(0,0){ 
			\includegraphics[width=0.495\linewidth,trim={2.2cm 0.cm 2.cm 1.cm},clip]{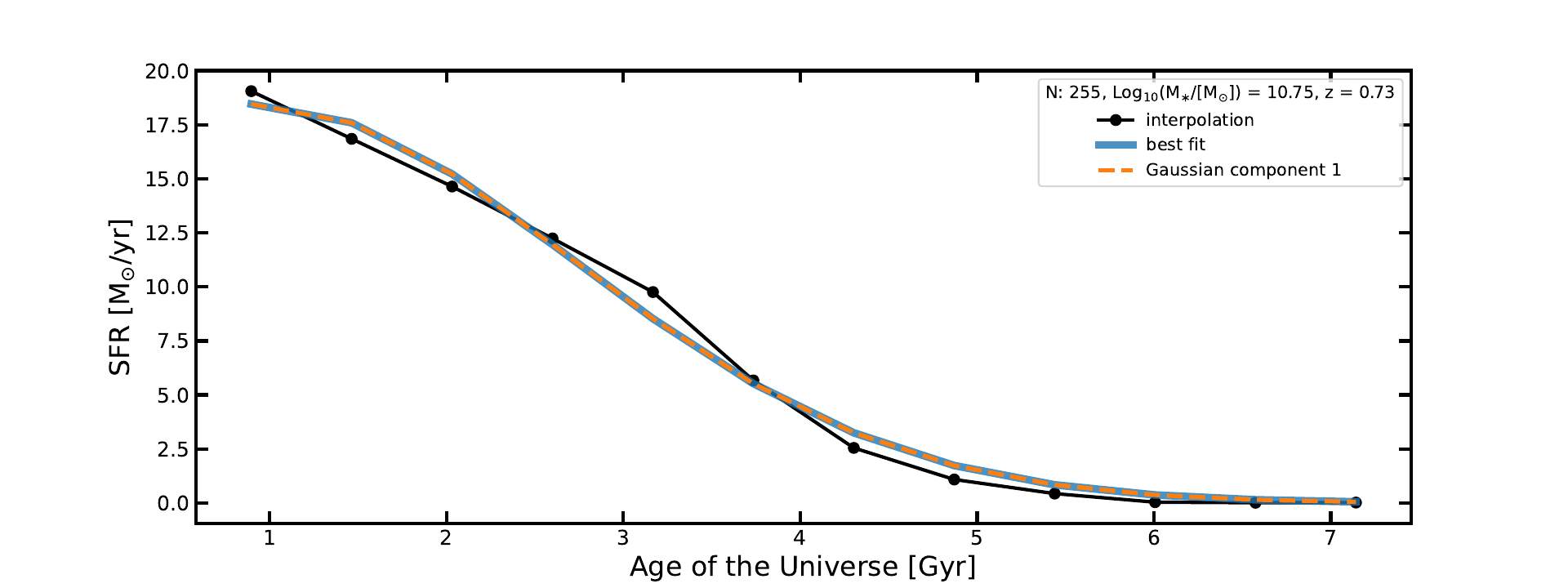}
			}
		\put(45,34){\tiny \scalebox{1.2}{\#98}}
	\end{picture}
    \\
    \begin{picture}(100,40)
		\put(0,0){ 
			\includegraphics[width=0.495\linewidth,trim={2.2cm 0.cm 2.cm 1.cm},clip]{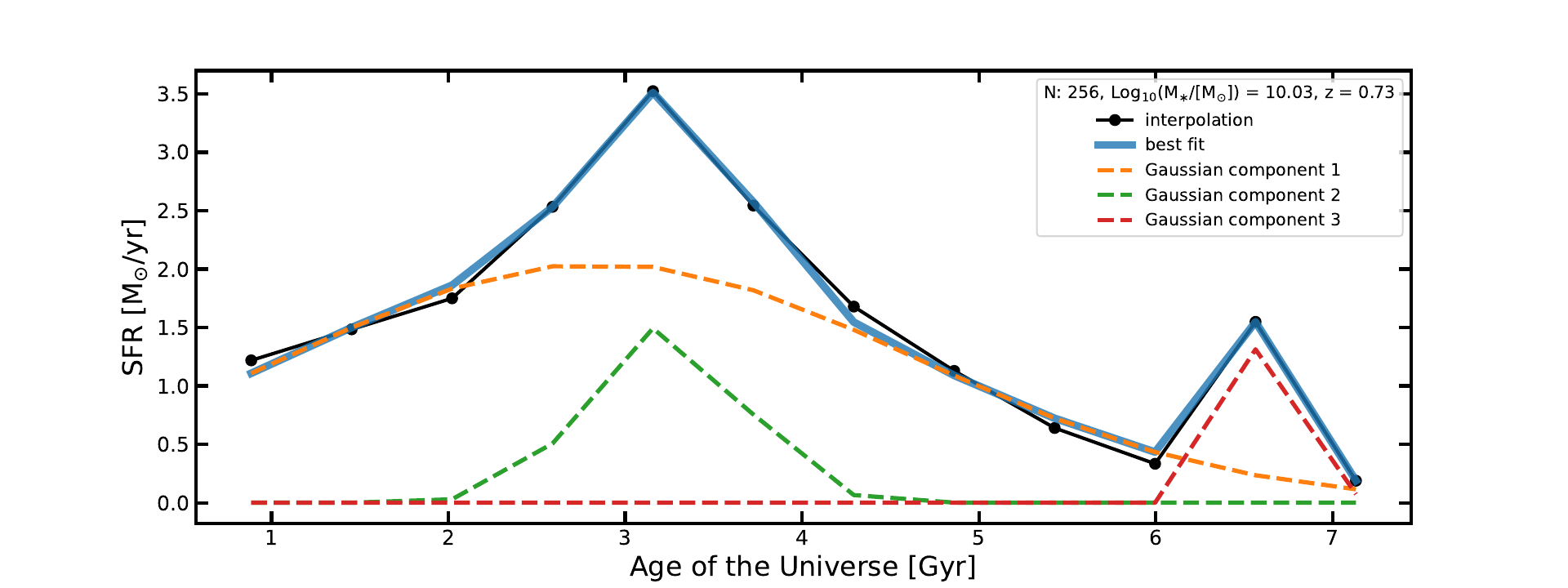}
			}
		\put(45,34){\tiny \scalebox{1.2}{\#105}}
	\end{picture}
    \hfill
    \begin{picture}(100,40)
		\put(0,0){ 
			\includegraphics[width=0.495\linewidth,trim={2.2cm 0.cm 2.cm 1.cm},clip]{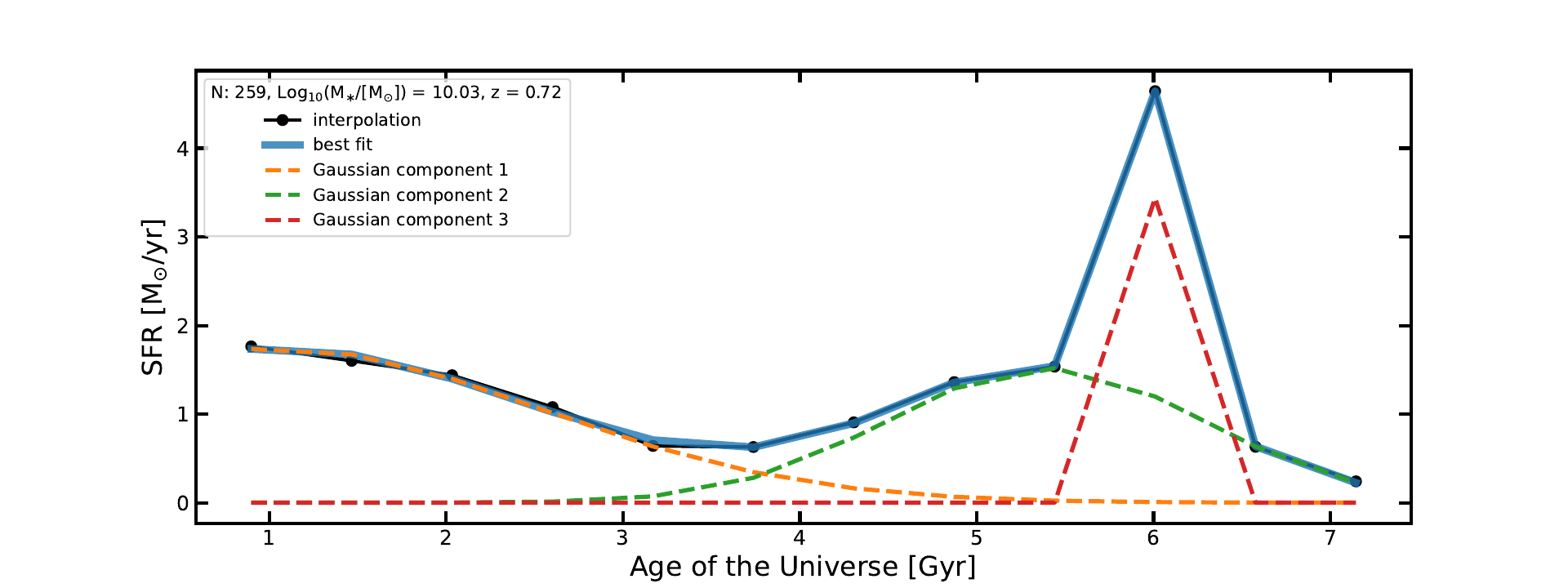}
			}
		\put(45,34){\tiny \scalebox{1.2}{\#119}}
	\end{picture}
    \\
    \begin{picture}(100,40)
		\put(0,0){ 
			\includegraphics[width=0.495\linewidth,trim={2.2cm 0.cm 2.cm 1.cm},clip]{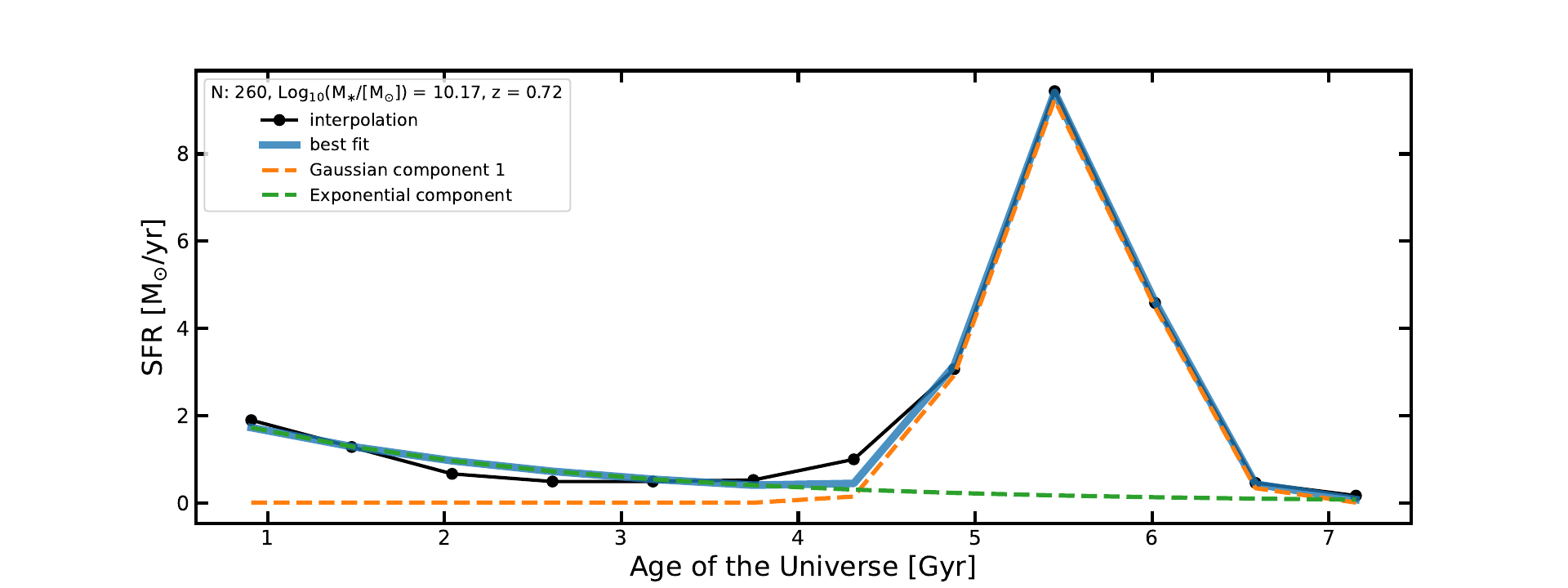}
			}
		\put(45,34){\tiny \scalebox{1.2}{\#131}}
	\end{picture}
    \hfill\hspace{0pt}
	\caption{
		Star formation histories for the CGr30 galaxies where available, from \citet{Munoz-Lopez2025}. For each panel, the black line shows the derived SFR in bins of 0.5 Gyr, the blue solid line the best-fit SFR, and the dashed lines the decomposition into different Gaussian or exponential components. The galaxy index $N$ in the \citet{Munoz-Lopez2025} is indicated in the legend. The four galaxies detected in CO (\# 71, \#105, \#119, \#131) are characterized by a recent burst of star formation, contrarily to the one that is not detected in CO (\# 98). 
        }
	\label{fig:sfh}
\end{figure*}

Figure~\ref{fig:sfh} shows the star formation histories (SFH) of the CGr30 galaxies where available, derived by \cite{Munoz-Lopez2025} within one effective radius. These SFH are obtained as byproducts of \texttt{pPXF} runs on the MUSE data, with the E-MILES stellar population synthesis templates and masked emission lines, see \citet{Munoz-Lopez2025} for more details. Interestingly, the SFHs of the four galaxies detected in CO are available and harbor a relatively recent burst of star formation: the three galaxies within the extended ionized gas structure (\#71, \#119, \#131) underwent a strong burst $1-2\rm~ Gyr$ ago, while the remaining galaxy (\#105) underwent a more limited one 0.5 Gyr ago. 
The SFH of one other CGr30 galaxy was derived by \cite{Munoz-Lopez2025}, namely \#98, and it contrasts with the previous ones by the absence of any recent burst.
Given the group virial radius and velocity dispersion ($R_{\rm 200}=669\rm ~kpc $, $\sigma_g=404~\rm km/s$, see Table~4 of \citealt{Epinat2024}), we estimate the group dynamical time ($R_{\rm 200}/\sigma_g$) to be $\sim 1.6~\rm Gyr$. 
The coincidence between the time since the last starburst for the three galaxies in the extended ionized gas structure and the group dynamical time could suggest that the starburst was triggered by the arrival of the galaxies within the group or a close encounter within it. Simulations and analytical models indeed show that tidal interactions can trigger starbursts via radial gas inflows, compressive tides and enhanced compressive turbulence leading to an excess of the star-forming dense gas \citep{Renaud2014, Renaud2015, Renaud2022}.  
However, these tidal interactions typically occur in the denser, inner regions of the group, where local interaction timescales are shorter than the global dynamical time. The estimated group dynamical time of 1.6~Gyr should thus be viewed as an upper limit for intragroup processing. The fact that the SFR peaks for the three galaxies within the extended ionized gas structure are comparable to this maximum timescale would then imply that the starbursts predate the galaxies' entry into the current group core, potentially pointing to pre-processing in smaller subgroups or interactions during the initial assembly phase. 
We also note that uncertainties in SFHs derived from full-spectrum fitting increase at older ages, such that broader components fitted at these epochs may reflect these expanding uncertainties rather than prolonged burst durations. 

\section{Depletion times}
\label{appendix:tdepl}

\begin{figure*}
	\centering
    \vspace{1cm}
    \begin{picture}(100,60)
    \put(-4,0){
		\includegraphics[width=0.49\linewidth,trim={0.cm 0.2cm 1cm 1.5cm},clip]{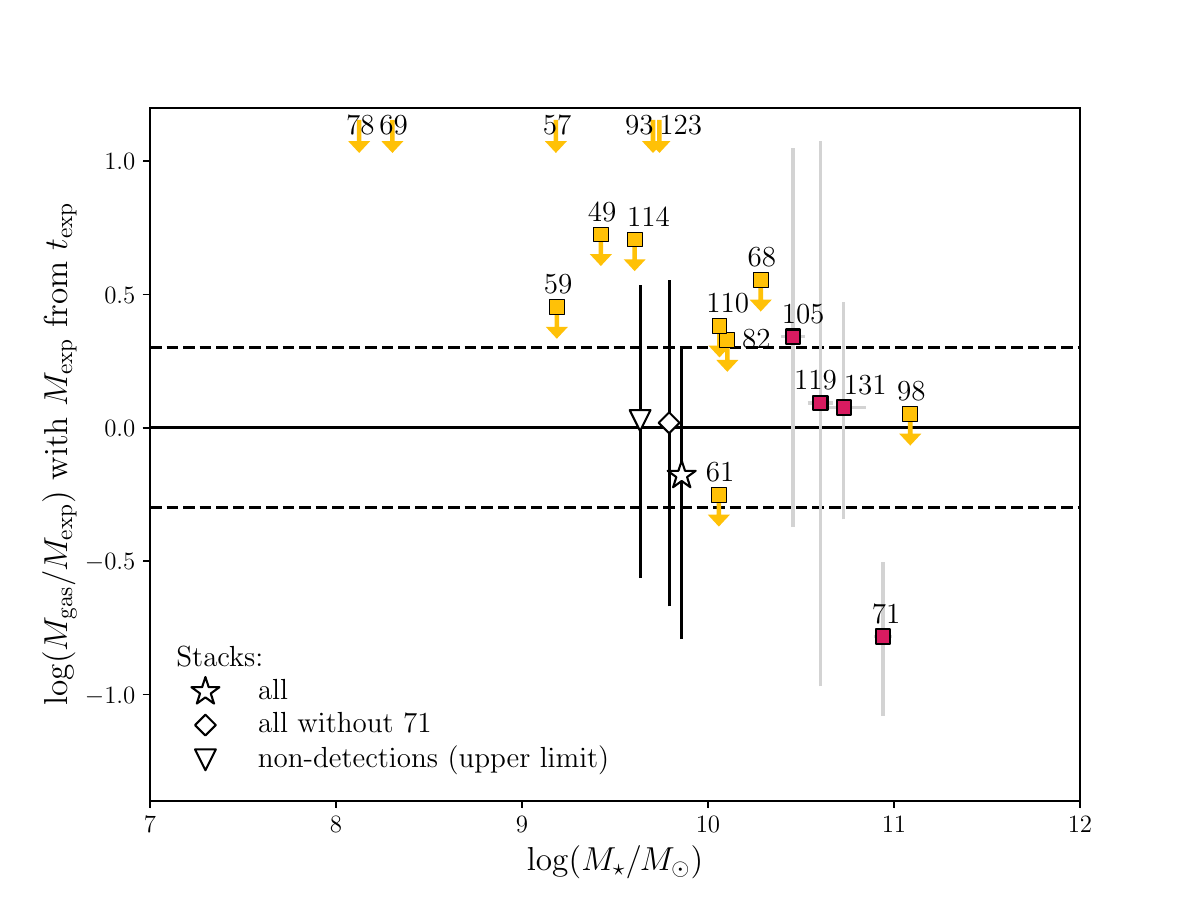}
	}
	\put(63,11){\tiny \scalebox{1.2}{SED-based SFRs}}
	\end{picture}
    \hfill
    \begin{picture}(100,60)
    \put(-4,0){
        \includegraphics[width=0.49\linewidth,trim={0.5cm 0.2cm 1cm 1.5cm},clip]{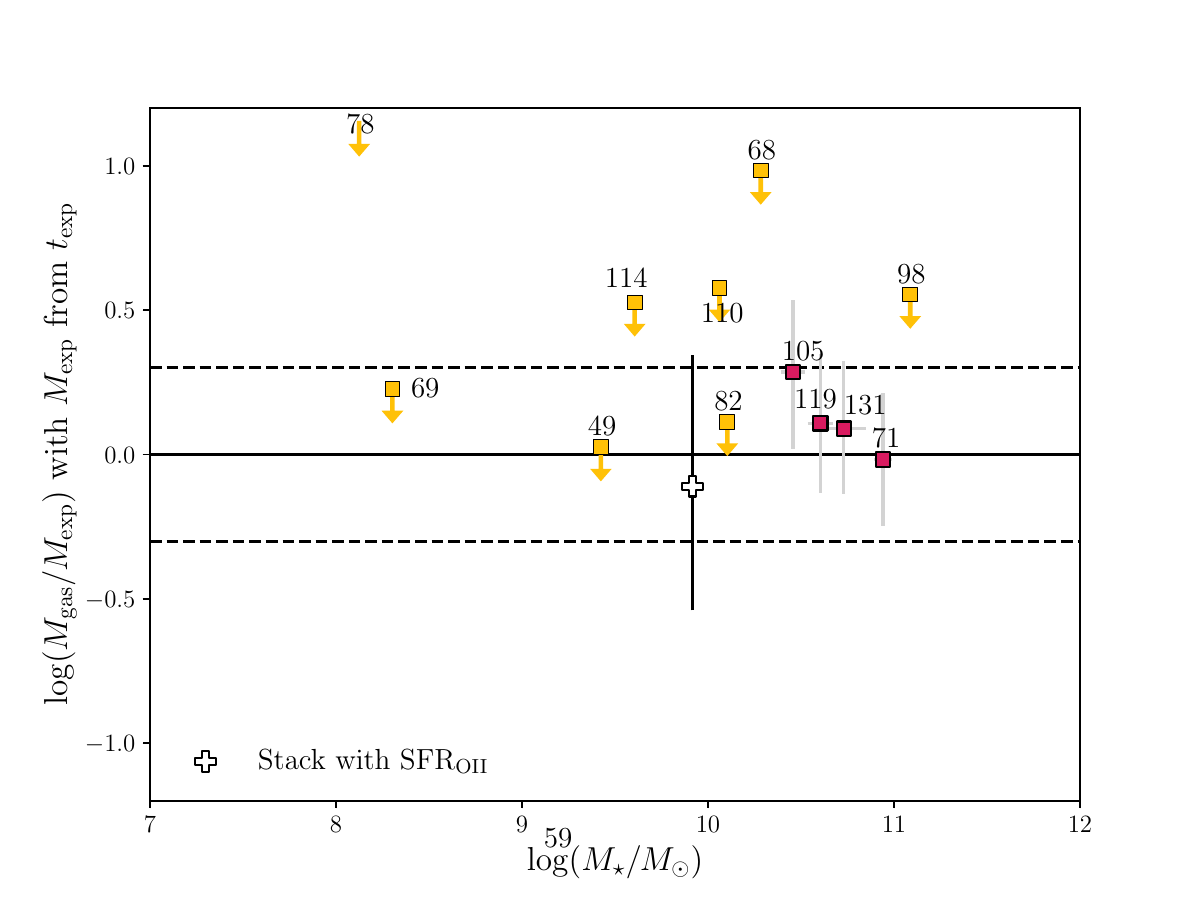}
	}
	\put(72,11){\tiny \scalebox{1.2}{\oii\ SFRs}}
	\end{picture}
	\caption{
		Molecular gas mass divided by that expected from the \citet{Tacconi2020} depletion time scaling relation as a function of the stellar mass, such that the ratio $M_{\rm gas}/M_{\rm exp}$ is a measure of $t_{\rm depl}/t_{\rm exp}$. 
        For SED-based SFRs \textit{(left panel)}, the stacked values for all galaxies of the sample amounts to $0.66\protect\substack{+1.35\protect\\-0.50}$ of the expected value from the scaling relation, without \#71 to $1.04\protect\substack{+2.55\protect\\-0.83}$, for the non-detections the upper limit to $1.07\protect\substack{+2.38\protect\\-0.79}$. 
        For \oii\ SFRs \textit{(right panel)}, the stacked measurement corresponds to $0.78\protect\substack{+1.44\protect\\-0.49}$ of the expected value from the scaling relation. 
        }
	\label{fig:mutdepl_vs_DMS_tdepl}
\end{figure*}

Figure~\ref{fig:mutdepl_vs_DMS} compares the molecular gas masses ($M_{\rm gas}$) with those expected ($M_{\rm exp}$) from the stellar masses and the \citet{Tacconi2020} molecular gas mass fraction scaling relation, such that the ratio $M_{\rm gas}/M_{\rm exp}$ is a measure of $\mu_{\rm gas}/\mu_{\rm exp}$. 
Alternatively, we can evaluate the expected molecular gas mass using the SFR and the depletion time scaling relation, which expresses the depletion time expected for typical star-forming galaxies as 
\begin{equation}
\log t_{\rm exp} = A + B\ \log(1+z) + C\  \delta {\rm MS} + D\ (\log M_\star-10.7)
\label{eq:Tacconi_tdepl}
\end{equation}
with $A=0.21$, $B=-0.98$, $C=-0.49$, $D=0.03$. In this case shown in Fig.~\ref{fig:mutdepl_vs_DMS_tdepl}, $M_{\rm gas}/M_{\rm exp}$ instead becomes a measure of $t_{\rm depl}/t_{\rm exp}$. 
As illustrated by the figure, the depletion times associated to both individual galaxies (except \#71 in the case of SED-based SFRs, which is an AGN) and the stacked measurements are in line with those of the typical star-forming galaxies underlying the \citet{Tacconi2020} scaling relations. 

\end{appendix}

\end{document}